\documentclass{amsart}

\usepackage{bbm}
\usepackage[lite]{amsrefs}
\usepackage{amssymb}
\usepackage[all,cmtip]{xy}
\usepackage{amsmath}
\usepackage{amsthm}
\usepackage{amsfonts}
\usepackage{hyperref}
\usepackage{enumerate}
\usepackage[titletoc]{appendix}
\usepackage{fancyhdr}
\usepackage[dvipsnames]{xcolor}
\usepackage{comment}
\usepackage{tikz-cd}
\usepackage{tikzpagenodes}
\usepackage{caption}
\usetikzlibrary[patterns]
\usepackage[margin = 1in]{geometry}
\usepackage{setspace}
\usepackage[T1]{fontenc}

\spacing{1.2}

\allowdisplaybreaks

 \newtheorem{thm}{Theorem}[section]

 \newtheorem{cor}[thm]{Corollary}
 \newtheorem{lem}[thm]{Lemma}
 \newtheorem{prop}[thm]{Proposition}
 
 \newtheorem{defn}[thm]{Definition}
  
  \newtheorem{defn-thm}[thm]{Definition-Theorem}
   \newtheorem{ex}[thm]{Example}
   
   \newtheorem{rem}[thm]{Remark}

 \newtheorem{notation}[thm]{Notation}

   \renewcommand\[{\begin{equation}}
\renewcommand\]{\end{equation}}
\numberwithin{equation}{section}

\usepackage{amsxtra}
\usepackage{graphicx}
\usepackage{bbm,newlfont}
\usepackage{amscd}
\usepackage{hyperref}
\usepackage[german,english]{babel}
\usepackage{cancel}
\usepackage{amsmath,amsthm,amssymb}
\usepackage{enumerate}
\usepackage{color}
\usepackage{varwidth}
\usepackage{mathrsfs}  
\usepackage{setspace}
\usepackage{collectbox}
\usepackage[utf8]{inputenc} 
\usepackage[T1]{fontenc}


\usepackage{graphicx}

\DeclareFontFamily{U}{mathx}{}
\DeclareFontShape{U}{mathx}{m}{n}{<-> mathx10}{}
\DeclareSymbolFont{mathx}{U}{mathx}{m}{n}
\DeclareMathAccent{\widehat}{0}{mathx}{"70}
\DeclareMathAccent{\widecheck}{0}{mathx}{"71}

\newcommand{\ZZ}{\mathbb{Z}}

\newcommand{\CC}{\mathbb{C}}

\newcommand{\nin}{\noindent}

\newcommand{\g}{\mathfrak{g}}

\newcommand{\n}{\mathfrak{n}}

\newcommand{\al}{\alpha}
\newcommand{\be}{\beta}

\newcommand{\vac}{\left| 0 \right>}

\newcommand{\ad}{\text{ad}}

\newcommand{\WW}{\mathcal{W}}

\begin{document}

\title{ Weak generators of W-algebras for type $A$ }
\author[M.H.Park]{Min Hee Park}
\address[M.H.Park]{Department of Mathematical Sciences, Seoul National University, GwanAkRo 1, Gwanak-Gu, Seoul 08826, Korea}
\email{mh6766@snu.ac.kr}

\author[U.R.Suh]{Uhi Rinn Suh}
\address[U.R.Suh]{Department of Mathematical Sciences and Research institute of Mathematics, Seoul National University, Gwanak-ro 1, Gwanak-gu, Seoul 08826, Korea}
\email{uhrisu1@snu.ac.kr}
\thanks{U.R. S. is supported by NRF Grant \#2022R1C1C1008698 and  Creative-Pioneering Researchers Program by Seoul National University}

\maketitle

\begin{abstract}
In this paper, we find weak generating sets for a classical 
W-algebra $\mathcal{W}^k(\g,f)$ when $\g=\mathfrak{sl}_N$ or $\mathfrak{sl}_{N_1|N_2}$. Furthermore, observing the relation between quantum and classical W-algebras, we further derive crucial information about the weak generating sets of quantum W-algebras at generic levels.

\end{abstract}


\setcounter{tocdepth}{-1}

\pagestyle{plain}

\section{Introduction}\label{Sec:Introduction}

A vertex algebra is first introduced by Borcherds \cite{Borch} as an underlying algebraic structure of 2 dimensional conformal field theory. In an algebraic point of view, it is equipped with infinitely many binary operations which are correlated by so-called the Borcherds identity. More precisely, a vertex algebra $V$ is endowed with the $n$-th product $(a,b) \mapsto a_{(n)}b$ for any $n\in \mathbb{Z}.$ For $n=-1$, this binary operation is called the normally ordered product. For $n<-1,$ we have an endomorphism $\partial$ on $V$ satisfying $:\partial^{-n-1}(a)b: \, = \,(-n-1)!\,  a_{(n)}b.$ The other products are from the singular part of operator product expansion (OPE). In this paper, we use $\lambda$-bracket notation introduced by Bakalov-Kac \cite{BK03}. In terms of $\lambda$-bracket, the singular OPE can be written as 
\begin{equation}\label{eq:lambda-bracket}
    [a{}_\lambda b]= \sum_{j\in \mathbb{Z}_+}\frac{\lambda^j}{j!} a_{(j)}b\in \mathbb{C}[\lambda]\otimes V,
\end{equation}
and a $\mathbb{C}[\partial]$-module endowed with a $\lambda$-bracket is called a Lie conformal algebra (LCA).
In the vertex algebra $V$ the subset $S$ is called a strong generating set of $V$ if every element in $V$ can be obtained from $S$ by the normally ordered product and the derivation $\partial$. We also define $S\subset V$ to be a weak generating set if it generates $V$ via all $n$-th product.

A Poisson vertex algebra (PVA) is a supercommutative, associative differential algebra equipped with a Lie conformal algebra (LCA) structure. Given a vertex algebra $V$, PVAs can be constructed in two distinct ways: through Li's filtration of $V$ introduced by Li \cite{Li04,Li05}, or via the quasi-classical limit of $V$
by De Sole-Kac \cite{DSK06}. Both approaches have been extensively used to investigate and derive structural properties of the original vertex algebra. 
 In the PVA theory, strong generating sets and weak generating sets are defined analogously to those in the theory of vertex algebras. In this paper, we examine the generators of a particular class of PVAs known as classical W-algebras, and derive properties of the weak generators of the corresponding vertex algebras referred to as quantum W-algebras.

Let $\g$ be a simple Lie algebra. A classical W-algebra $\mathcal{W}^k(\g)$ for $k\in \CC$ is first introduced by Drinfeld-Sokolov \cite{DS85} in the context of integrable systems as a Hamiltonian reduction of the classical affine PVA $\mathcal{V}^k(\g).$ Their results are generalized to a larger family of classical W-algebras $\mathcal{W}^k(\g,f)$ by De Sole-Kac-Valeri, where $f$ is any nilpotent element $f\in \g$ (see \cite{DSKV13,DSKV15,DSKV16b} and references therein). In addition, the same authors \cite{DSKV16c} described the complete OPE structures of these PVAs. When $\g$ is a basic Lie superalgebra, the OPE structure of $\mathcal{W}^k(\g,f)$ is written by the second author \cite{Suh16,Suh20} and its integrability has been studied in \cite{Suh18}.
Notice that $\mathcal{W}^k(\g)=\mathcal{W}^k(\g,f_{\text{prin}})$ and $W^k(\g)=W^k(\g,f_{\text{prin}})$ for the principal nilpotent $f_{\text{prin}}.$ 

On the other hand, a quantum W-algebra $W^k(\g)$ of level $k\in \mathbb{C}$ is constructed by Feigin-Frenkel \cite{FF90} by so-called BRST complex, and Kac-Roan-Wakimoto \cite{KRW04} introduced quantum W-algebras $W^k(\g,f)$ for any basic Lie superalgebra $\g$ and even nilpotent element $f.$ It is known by De Sole-Kac \cite{DSK06} that $W^k(\g,f)$ is a quantization of $\mathcal{W}^k(\g,f)$, in the sense that the quasi-calssical limit of the BRST complex gives rise to $\mathcal{W}^k(\g,f)$. If one considers Li's filtration of the BRST complex, the resulting PVA is the classical W-algebra $\mathcal{W}^{k=0}(\g,f)$. This particular classical W-algebra has been studied in relation to the geometric approach to W-algebras (see for example \cite{Ara12,Ara17}). However, for the purpose of this paper, the level $k\neq 0$ classical W-algebra is more suitable. 

In structural theories, strong generating sets of W-algebras are studied well \cite{DSK06,DSKV16c,Suh20}. For classical W-algebra $\mathcal{W}^k(\g,f),$ one can find a unique differential algebra isomorphism $\omega$ between the algebra generated by the centralizer $\g^f$ of $f$ and $\mathcal{W}^k(\g,f).$ This result is natural since the finitizaion of $\mathcal{W}^k(\g,f)$ is the Poisson structure of a Slodowy slice and it also implies that $\omega(a)$ for $a\in \g^f$ strongly generates $\mathcal{W}^k(\g,f).$ Moreover, the Poisson brackets between the generators are also well-understood. The explicit formulas (Theorem \ref{thm:main lemma}) are written in \cite{DSKV16c,Suh20} and it can alternatively be computed by the Dirac reduction method \cite{LSS23}. For the quantum W-algebra $W^k(\g,f)$, De Sole and Kac \cite{DSK06} showed it is a universal enveloping algebra of a certain nonlinear LCA and hence there is a strong generating set consisting of $\text{dim}\g^f$ elements. 

The strong generating sets of W-algebras are known to be minimal when viewed as differential algebras, but they are not minimal when considered as vertex algebras. In other words, there are weak generating sets of $W^k(\g,f)$ whose orders are strictly less than $\text{dim}(\g^f).$
For example, the principal W-algebra $W^k(\mathfrak{sl}_n)$ of non-critical level $k$ is weakly generated by weight 2 and 3 fields \cite{ALY19,L21}.
It can be easily derived from the property of $W_{1+\infty}$ algebra in \cite{FKRW95} that a weight $n$ or $n-1$ field of $W^k(\mathfrak{sl}_n)$ weakly generates the whole algebra at generic levels. In addition, weak generating sets of non-super classical types and $A(n+1|n)$ type principal W-algebras are also investigated in \cite{CKL24} and \cite{CKLSS25}. In these papers, the authors described how to get all other strong generators, starting from the given weak generators. In this paper, we deal with type $A$ classical and quantum W-(super)algebras associated with any even nilpotent element $f.$ More precisely, we show the following Theorem.

\begin{thm}[Theorem \ref{big thm} and Theorem \ref{big thm small}] \label{thm:main}
Let $\g=\mathfrak{sl}_N$ and  $f$ be the nilpotent with the partition $(m_1, m_2, \cdots, m_d)$, where $m_1\geq m_2\geq \cdots \geq m_d$.
For each $(i,j)$ for $i, j=1,\cdots, d,$
denote by $q^{ij}_{m}\in \g^f$ a weight $m$ element located in the $(i,j)$-block matrix $B^{(i,j)}$ defined in \eqref{eq:block matrix}.
Recall the differential algebra isomorphism $\omega:S
(\mathbb{C}[\partial]\otimes \g^f)\to \mathcal{W}^k(\g,f)$. Then each of following sets weakly generate $\mathcal{W}^k(\g,f)$ for $k\neq0$:
\begin{equation*}
    \mathcal{C}^f_{\text{big}}= \{\omega(b)| b\in \mathcal{B}^f_{\text{big}}\}, \quad  \mathcal{C}^f_{\text{sm}}= \{\omega(b)| b\in \mathcal{B}^f_{\text{sm}}\},
\end{equation*}
where $\mathcal{B}^f_{\text{big}}$ and $\mathcal{B}^f_{\text{sm}}$ are defined as follows.
\begin{enumerate}[(1)]
\item If $m_1>m_2$, then 
\begin{align}
   &  \mathcal{B}^f_{\text{big}}= \Big\{ q_{\frac{m_{i}+m_{i+1}}{2}}^{(i,i+1)}\,,\,  q_{\frac{m_{i}+m_{i+1}}{2}}^{(i+1, i)} \, \Big| \, 1\leq i\leq d-1 \Big\}, \nonumber \\
   & \mathcal{B}^f_{\text{sm}}= \Big\{ q_{\,3}^{(1,1)}\,,\, q_{\frac{m_i-m_{i+1}}{2}+1}^{(i,i+1)}\,,\, q_{\frac{m_i-m_{i+1}}{2}+1}^{(i+1,i)} \, \Big|\, 1\leq i\leq d-1\Big\}, \quad (m_1>2). \nonumber
\end{align}
 Note that if $m_1=2$, then $q_{\,3}^{(1,1)}$ does not exist and, in this case, we get a weak generating set from   $\mathcal{C}^f_{\text{sm}}$ by including  $\omega\big(q_{\,2}^{(1,1)}\big)$ instead of $\omega\big(q_{\,3}^{(1,1)}\big).$
\item If $m_1=m_2$, then 
\begin{align}
&    \mathcal{B}^f_{\text{big}}= \Big\{ q_{\,{m_1}-1}^{(2, 1)}\,,\,  q_{\frac{m_{i}+m_{i+1}}{2}}^{(i,i+1)}\,,\,   q_{\frac{m_{j}+m_{j+1}}{2}}^{(j+1, j)}\, \Big| \   1\leq i\leq d-1, \ 2\leq j\leq d-1 \ \Big\}, \nonumber \\
& 
\mathcal{B}^f_{\text{sm}}= \Big\{ q_{\,3}^{(1,1)}\,,\, q_{\,2}^{(2,1)}\,,\, q_{\frac{m_i-m_{i+1}}{2}+1}^{(i,i+1)}\,,\, q_{\frac{m_j-m_{j+1}}{2}+1}^{(j+1,j)}\, \Big| \, 1\leq i\leq d-1, \ 2\leq j\leq d-1 \, \Big\}. \nonumber
\end{align}
As in (1), $\mathcal{C}^f_{\text{sm}}$ is a weak generating set when $m_1\geq 3$. If $m_1=2$, then  we get a weak generating set from   $\mathcal{C}^f_{\text{sm}}$ by including  $\omega\big(q_{\,2}^{(1,1)}\big)$ instead of $\omega\big(q_{\,3}^{(1,1)}\big).$
\end{enumerate}
\end{thm}

In Theorem \ref{thm:main}, the set $\mathcal{C}^f_{\text{big}}$ consists of homogeneous elements with bigger conformal weights and  $\mathcal{C}^f_{\text{sm}}$  consists of homogeneous elements with smaller conformal weights.
To prove the theorem, the explicit form of Poisson brackets on $\mathcal{W}^k(\g,f)$ plays a crucially role. Moreover, by analogous computations, the same result extends to the classical W-superalgebras: 

\begin{thm}[Theorem \ref{big thm super} and Theorem \ref{big thm small super}] \label{thm:main2}
Let $\g=\mathfrak{sl}_{N_1|N_2}$ with $N_1\neq N_2$ and $f$ be the nilpotent with the partition $(m_1, m_2, \cdots, m_{d_1})$ of $N_1$ and the partition $(m_{d_1+1}, m_{d_1+2}, \cdots, m_{d_1+d_2})$ of $N_2$, where $m_1\geq m_2\geq \cdots \geq m_{d_1}$, $m_{d_1+1}\geq m_{d_1+2}\geq \cdots \geq m_d$ and $d=d_1+d_2$.
We denote by $m_l$ the larger of $m_1$ and $m_{d_1+1}$. 
Then each of following sets weakly generate $\mathcal{W}^k(\g,f)$ for $k\neq0$:
\begin{equation*}
    \mathcal{C}^f_{\text{big}}= \{\omega(b)| b\in \mathcal{B}^f_{\text{big}}\}, \quad  \mathcal{C}^f_{\text{sm}}= \{\omega(b)| b\in \mathcal{B}^f_{\text{sm}}\},
\end{equation*}
where $\mathcal{B}^f_{\text{big}}$ and $\mathcal{B}^f_{\text{sm}}$ are defined as follows.
The set $\mathcal{B}^f_{\text{big}}$ coincides with that in Theorem \ref{thm:main}. The set $\mathcal{B}^f_{\text{sm}}$ has the same form as  in the corresponding theorem, except that $q_{\,2}^{(1,1)}$ and $q_{\,3}^{(1,1)}$ are replaced by  $q_{\,2}^{(l,l)}$ and $q_{\,3}^{(l,l)}$, respectively, and the term $m_i-m_{i+1}$ in its elements is replaced by $\lvert m_i-m_{i+1}\rvert$.
\end{thm}

In order to look for weak generating sets of the quantum W-algebra $W^k(\g,f),$ recall the vertex algebra $W^k(\g,f)_{\epsilon}$ over the ring $\mathbb{C}[\epsilon,\epsilon^{-1}],$ which was introduced by De Sole-Kac \cite{DSK06}. They showed that $W^k(\g,f)_{\epsilon}|_{\epsilon=0}\simeq \mathcal{W}^k(\g,f)$ and $W^k(\g,f)_{\epsilon}|_{\epsilon=1}\simeq W^{k}(\g,f)$. Using the property, we show the following theorem.

\begin{thm} [Theorem \ref{thm:weak generator-quantum case}] \label{thm:main3}
Let $\g=\mathfrak{sl}_N$ or $\mathfrak{sl}_{N_1|N_2}$ and let $f$ be the nilpotent element of $\g$ in Theorem \ref{thm:main} or \ref{thm:main2}. Each of the set 
\[\{ J^{\epsilon=1}_{\{b\}}\, |\, b\in \mathcal{B}^f_{\text{big}}\},\  \{ J^{\epsilon=1}_{\{b\}}\, |\, b\in \mathcal{B}^f_{\text{sm}}\}\subset W^{k}(\g,f)\]
weakly generates $W^{k}(\g,f)$ for generic values of $k \in \mathbb{C}$. 
\end{thm}

Theorem \ref{thm:main3} can be interpreted as a weak generation property of $W^k(\g,f)$, analogous to the strong generation property established in \cite{DSK06}. However, the OPE relations between the weak generators and the precise values of level at which these weak generating sets exist remain open questions. Furthermore, we will work on the sufficient condition to determine the OPE relations of W-algebras, together with the weak generators.

\section{Preliminaries} \label{sec:prelim}
In this section, we review the structure of both the classical and quantum affine W-algebras. 
After describing the generators and relations of the classical affine W-algebra $\mathcal{W}^k(\g, f)$ introduced in \cites{DSKV16c, Suh20}, we examine the correspondence between the classical and quantum affine W-algebras. 
We first recall basic notations related to Poisson vertex algebras and classical W-algebras. 

\subsection{Poisson vertex algebras}
Let $V=V_{\bar{0}}\oplus V_{\bar{1}}$ be any vector superspace. 
If $a\in V_{\bar{0}}$ (resp. $a\in V_{\bar{1}}$), the parity of $a$ is defined by $p(a)=0$ (resp. $p(a)=1$).
 If a linear operator $\phi\colon V\to V$ satisfies $\phi(V_{\bar{i}})\subset V_{\bar{i}}$ (resp. $\phi(V_{\bar{i}})\subset V_{\overline{i+1}}$)  for  $i\in\{0, 1\}$, denote its parity by $p(\phi)=0$ (resp. $p(\phi)=1$).
 $V$ is called a superalgebra if it satisfies $V_{\bar{i}}V_{\bar{j}}\subset V_{\overline{i+j}}$, for $i, j\in\{0, 1\}$.
 If there is a linear operator $d$ on $V$ satisfying
$d(ab)=d(a)b+(-1)^{p(a)p(d)}a\, d(b)$
for $a, b\in V$, then $V$ is called a differential algebra with the derivation $d$.

Let $R=R_{\bar{0}}\oplus R_{\bar{1}}$ be a $\CC[\partial]$-module with an even operator $\partial\colon R\to R$. 
Consider the even indeterminate $\lambda$ and $\CC[\partial]$-module $\CC[\lambda]\otimes R$, defined by the following formula,
\begin{align*}
\partial(\lambda^n\otimes a)=\lambda^n\otimes\partial a,
\end{align*}
for $a\in R$. 
Throughout this paper, we omit $\otimes$ in elements of $\CC[\lambda]\otimes R$ and assume that $a$ is a homogeneous element unless otherwise stated. 
Consider a $\CC[\partial]$-module $R$ endowed with a parity preserving bilinear $\lambda$-bracket $[\cdot _\lambda \cdot]:R\otimes R\to \CC[\lambda]\otimes R$ satisfies the following axioms:
\begin{enumerate}[]
\item (sesquilinearity) $[\partial a _\lambda b]=-\lambda[a _\lambda b]$ and $[a _\lambda \partial b]=(\partial+\lambda)[a _\lambda b],$
\item (skewsymmetry) $[a_\lambda b]=-(-1)^{p(a)p(b)}[b _{-\lambda-\partial} a],$
\item (Jacobi identity) $[a_\lambda [b _\mu c]]=[[a_\lambda b] _{\lambda+\mu}c]+(-1)^{p(a)p(b)}[b_ \mu [a _\lambda c]]$ in $\CC[\lambda, \mu]\otimes R$.
\end{enumerate} 
Then we call $R$ \textit{a Lie (super)conformal algebra (LCA)}.
Let us denote 
 \begin{align*}
 [a _\lambda b]=\sum_{n\in \mathbb{Z}_+}\frac{\lambda^n}{n!}a_{(n)}b,
 \end{align*}
 for $a, b, a_{(n)}b\in R$. Then the RHS of the skew-symmetry is rewritten as 
$ [b{}_{-\lambda-\partial} a]=\sum_{n\in \mathbb{Z}_+}\frac{(-\partial-\lambda)^n}{n!}b_{(n)}a$ and terms in Jacobi identity are 
        \begin{align*} [a{}_\lambda[b{}_\mu c]]=\sum_{m,n\in \mathbb{Z}_+}\frac{\lambda^n\mu^m}{n!m!}a_{(n)}(b_{(m)}c), \quad  [[a{}_\lambda b]{} _{\lambda+\mu} c]=\sum_{m,n\in \mathbb{Z}_+}\frac{(\mu+\lambda)^m\lambda^n}{m!n!}(a_{(n)}b)_{(m)}c.\end{align*}

Now the Poisson vertex algebra, which is an algebraic structure we mainly concern in the paper can be introduced.
\begin{defn}  \label{def:PVA}
    A \textit{Poisson vertex algebra (PVA)} is a quadruple $(\mathcal{P}, \partial, \{\ _\lambda \ \}, \cdot)$ satisfying the following three properties:
    \begin{enumerate}[(i)]
    \item $(\mathcal{P}, \partial, \{\ _\lambda \ \})$ is a LCA,
    \item $(\mathcal{P}, \partial, \cdot)$ is a unital supercommutative associative differential algebra, and
    \item (Leibniz rule) $\{a _\lambda bc\}=\{a _\lambda b\}c+(-1)^{p(a)p(b)}b\{a _\lambda c\}$ for $a, b, c\in\mathcal{P}$.
    \end{enumerate}
\end{defn}

By definition, a PVA has a LCA structure. Conversely, from a LCA, one can construct a PVA (\cite{BDSK09}).
For a vector superspace $V=V_{\bar{0}}\oplus V_{\bar{1}}$ and denote by $S(V)=S(V_{\bar{0}}) \otimes \bigwedge (V_{\bar{1}})$ the supersymmetric algebra of $V$. Then the following properties hold. When a LCA $R$ is given, the supersymmetric algebra $S(R)$ generated by $R$ is a PVA endowed with the $\lambda$-bracket defined by the $\lambda$-bracket on $R$ and the Leibniz rule. Moreover, if $\mathcal{P}$ is a differential algebra generated by a vector superspace $V$ then the linear map $V\otimes V\to \mathcal{P}[\lambda]$ satisfying the skew-symmetry and Jacobi identity can be extended to a PVA $\lambda$-bracket on $\mathcal{P}$ via the sesquilinearity and the Leibniz rule. 

Many of interesting PVAs have a special element $L$ called a \textit{conformal vector} of $\mathcal{P}$. This element satisfies  (i) $
[L\ _\lambda\ L]=(\partial+2\lambda)L+\frac{c}{12}
\lambda^3$
for some $c\in\CC$, (ii) $L_{(0)}=\partial$, (iii) $L_{(1)}$ is a diagonalizable operator on $\mathcal{P}$. The eigenvalue of a homogeneous element $a\in \mathcal{P}$ is denoted by $\Delta_a$ and called the {\it conformal weight} of $a$. The conformal weight has the properties 
\begin{equation*}
    \Delta_1=0, \quad \Delta_{a_{(n)}b}=\Delta_a+\Delta_b-n-1, \quad \Delta_{\partial a}=\Delta_a+1.
\end{equation*}

\vskip 2mm

Before we close this section, we introduce two distinct types of generating sets of a PVA called strong and weak generating sets, which play a crucial role in this paper. 

\begin{defn}
Let $\mathcal{P}$ be a PVA. A subset $\widetilde{S}\subset \mathcal{P}$ is said to strongly generates $\mathcal{P}$ if it generates $\mathcal{P}$ as a differential algebra generated by $\widetilde{S}$. A subset $S$ is said to weakly generates $\mathcal{P}$ if it generates $\mathcal{P}$ as a PVA, i.e. via all $n$-th products for $n\in \mathbb{Z}.$ By definition, every strong generating set is also a weak generating set, though the converse does not hold.
\end{defn}

\subsection{Classical W-algebras} \label{subsec:classical W}
Now, we introduce the definition of classical affine W-algebra $\mathcal{W}^k(\g, f)$.
Let $\g$ be a basic finite Lie superalgebra with an $\mathfrak{sl}_2$-triple $(f, h=2x, e)$. 
Suppose that $\g$ has a normalized nondegenerate supersymmetric invariant even bilinear form $(\ | \ )_{nor}$ 
 on $\g$ such that $(\theta|\theta)_{nor}=2$ for an even highest root $\theta$.
Then $\g$ decomposes into a direct sum of $\ad x$-eigenspaces, denoted as $\g=\bigoplus_{i\in\frac{\ZZ}{2}}\g_i$.
Let $\mathcal{V}^k(\g)=S(\CC[\partial]\otimes \g)$ be a PVA with the $\lambda$-bracket 
\begin{align}\label{bracket of v^k(g)}
\{a _\lambda b\}=[a, b]+k\lambda(a| b)_{nor},
\end{align}
for $a,b\in \g.$ For the rest of paper, we consider the re-normalized bilinear form $(\ | \ )$ on $\g$ so that $(e|f)=\frac{1}{2}(h|h)=1$ and denote by $\tilde{k}$ the constant satisfying $k(a|b)_{nor}=\tilde{k}(a|b).$
Let us consider the subspaces of $\g$:
\begin{align*}
\mathfrak{p}=\mathop{\bigoplus}_{i<1}\g_i, \quad \mathfrak{m}=\mathop{\bigoplus}\limits_{i\geq1}\g_i, \quad   
\mathfrak{n}=\mathop{\bigoplus}\limits_{i>0}\g_i,
\end{align*}
  and denote by $\mathcal{V}(\mathfrak{p}):=S(\CC[\partial]\otimes\mathfrak{p})$ the differential subalgebra of $\mathcal{V}^k(\g)$  generated by $\mathfrak{p}$. Define the differential algebra homomorphism $\rho:\mathcal{V}(\g)\to\mathcal{V}(\mathfrak{p})$ on generators by 
 $
\rho(a)=\pi_\mathfrak{p}(a)+(f|a), \ a\in\g,$
where $\pi_\mathfrak{p}:\g\to\mathfrak{p}$ is the projection with the kernel $\mathfrak{m}$. 
\begin{defn} \label{def:classical W}
The classical affine W-algebra is defined by the following differential algebra, 
\begin{align*}
\mathcal{W}^k(\g, f)=\{A\in\mathcal{V}(\mathfrak{p})\mid\rho\{n_\lambda A\}=0 \mathrm{\ for \ all \ } n\in\mathfrak{n}\}
\end{align*}
equipped with the PVA $\lambda$-bracket 
$
\{A_\lambda B\}=\rho\{A_\lambda B\}$
for $A,B\in\mathcal{W}^k(\g, f)$.
\end{defn}

Let $k\neq 0$. It is known that $\mathcal{W}^k(\g,f)$ has a conformal vector which is induced from the conformal vector $L=L_{\text{sug}}+\partial x$ of $\mathcal{V}^k(\g)$, where $L_{\text{sug}}$ is the sugawara operator. Hence each element $a\in \g_i \subset \mathcal{V}^k(\g)$ has the conformal weight $\Delta_a=1-i.$ The conformal weight of an element in $\mathcal{W}^k(\g,f)$ is naturally induced from that of the corresponding element in $\mathcal{V}^k(\g)$. In addition, it is known by \cites{DSKV13, Suh20} that $\mathcal{W}^k(\g,f)$ is isomorphic to $S(\CC[\partial]\otimes \g^f)$  as a differential algebra, where $\g^f$ denotes the centralizer of $f$ in $\g$.
Precisely, there is the unique differential algebra isomorphism 
\begin{equation} \label{eq:omega} \omega:S(\CC[\partial]\otimes\g^f)\xrightarrow{\sim}\mathcal{W}^k(\g, f),
\end{equation}
satisfying the following properties \cites{DSKV16c,Suh20}:
\begin{enumerate}[(i)]
\item For $a\in\g^f$, the conformal weight of $\omega(a)$ is equal to the conformal weight of $a$,
\item In $\omega(a)$, the linear term without total derivative is equal to $a$.
\end{enumerate}
Hence the $\mathcal{W}^k(\g,f)$ is {\it strongly generated } by $\omega(q_j)$ for $j\in I^f$, where $\{q_j|j\in I^f\}$ is a basis of $\g^f$.  
Note that the conformal vector of $\mathcal{W}^k(g,f)$ can be expressed via the isomorphism $\omega$ by 
$
L=\omega(f)+\frac{1}{2}\sum_{j\in J^f_0}\omega(q_j)\omega(q_j^*),
$
where $\{q_j|j\in J^f_0\}$ and $\{q_j^*|j\in J^f_0\}$ are bases of $\g^f\cap \g_0$ satisfying $(q_i^*|q_j)=\delta_{ij}.$

Now let us recall the explicit formula of the Poisson $\lambda$-bracket of $\mathcal{W}^k(\g,f).$
By the $\mathfrak{sl}_2$ representation theory, the centralizers $\g^f=\ker(\ad f)$ and $\g^e=\ker(\ad e)$ of $f$ and $e$ are nondegenerately paired by $(\cdot| \cdot)$. 
To briefly recall generators and relations between generators of $\mathcal{W}(\g, f)$ introduced in \cites{DSKV16c, Suh20}, 
take bases $\mathcal{B}^f:=\{q_j |\,  j\in J^f\}$ and $\mathcal{B}^e:=\{q_j^* |\, j\in J^f\}$ of $\g^f$ and $\g^e$, respectively, whose elements are all homogeneous and  $\ad x$-eigenvectors. Moreover, we assume  
$(q^*_i | q_j)=\delta_{i, j}.$
For convenience, we denote $p(j)=p(q_j)$ and $s(j)=s(q_j)=(-1)^{p(j)}$. 
Also, for $j\in J^f$, let $\delta(j)$ be the $\ad x$-eigenvalue of $q_j^*$. As extended bases of $\g$ from $\mathcal{B}^f$ and $\mathcal{B}^e$, respectively, we get dual bases 
\begin{equation} \label{eq:basis of g}
    \begin{aligned}
        & \mathcal{B}=\{\, q_{j}^*[n]:=(\ad f)^n q_j^* \, |\,  j\in J^f, 0\leq n\leq2\delta(j)\, \},\\
        & \mathcal{B}^{*}=\Big\{q_{j}[n]:=\frac{(-1)^n}{(n!)^2{2\delta(j) \choose n}} (\ad e)^n q_j \, \Big|\,  j\in J^f, 0\leq n\leq2\delta(j)\Big\}
    \end{aligned}
\end{equation}
of $\g$ and thus 
\begin{equation*}
    (\, q_i^*[m]\, |\, q_j[n]\, )=\delta_{ij}\delta_{mn}.
\end{equation*}
In the set  $\bigsqcup_{l\in \mathbb{Z}/2}J_l^f$ of indices where $J_l^f:=\{(j, n)\in J^f\times\ZZ_{+}\mid q_j[n]\in\g_l \ \text{or} \ \ q_j^*[n]\in\g_{-l}\}$, we consider a partial order defined as follows: 
\begin{itemize}
\item $(j_t, n_t)\prec(j_{t+1}, n_{t+1})$ if and only if $\begin{cases}\alpha_{t+1}-\alpha_t\geq1 \ \mathrm{where} \\ (j_t, n_t)\in J_{\alpha_t}^f \ \mathrm{and} \ (j_{t+1}, n_{t+1})\in J_{\alpha_{t+1}}^f.\end{cases}$    
\item $(j, n)\prec l$ (resp. $l\prec (j, n))$ if and only if $(j,n)\in J^f_\alpha$ for $\alpha\leq l-1$ (resp. $l\leq\alpha-1)$. 
\end{itemize}
The following theorem shows the explicit formula for the $\lambda$-brackets between strong generators of $\mathcal{W}^k(\g,f)$.

\begin{thm}\cites{DSKV16c, Suh20} \label{thm:main lemma}
Let us denote $\g^f_t:=\g^f \cap \g_t$ and recall the isomorphism $\omega$ in \eqref{eq:omega}. 
For $a\in \g^f_{-t_1}$ and $b\in\g^f_{-t_2}$, the $\lambda$-bracket between $\omega(a)$ and $\omega(b)$ in $\mathcal{W}^k(\g,f)$ is as follows:
\begin{equation*}
\begin{aligned}
&\{\omega(a)_\lambda \omega(b)\}\nonumber =\omega([a, b])+\tilde{k}\lambda(a|b)
\\&-s(a, b)\sum\limits_{p\in\ZZ_{+}}\sum\limits_{\substack{{-t_2-1\prec(j_0, n_0)\prec\cdots} \\ {\cdots\prec(j_{p}, n_{p})\prec t_1}}}\bigg(\prod_{r=0}^{p} s(j_r)\bigg)\Big(\omega\big(\, [\, b\, ,\,  q_{j_0}^*[n_0]\, ]^\sharp\, \big)-(\, b\mid q_{j_0}^*[n_0]\, ) \, \tilde{k}(\lambda+\partial)\Big)
\\& \hskip 2cm\left[\prod_{t=1}^p \Big(\omega \big(\, \big[\, q_{j_{t-1}}[{n_{t-1}+1}]\, ,\,  q_{j_t}^*[n_t]\, \big]^\sharp\, \big)-( \, q_{j_{t-1}}[{n_{t-1}+1}]\, \big| \,  q_{j_t}^*[n_t]\, )\, \tilde{k}(\lambda+\partial)\Big)\right] 
\\&  \hskip 2cm \Big(\, \omega\big(\, [\, q_{j_p}[n_p+1]\, ,\,  a\, ]^\sharp\, \big)-(\, q_{j_p}[n_p+1]\, | \,  a\, )\, \tilde{k}\lambda \Big).
\end{aligned}
\end{equation*}
Here $g^\sharp$ for $g\in \g$ denotes the image of projection map $\pi_f:\g\to \g^f$ corresponding to the decomposition $\g=\g^f\oplus [\g,e].$
\end{thm}

We know that $\mathcal{W}^k(\g,f)$ is the algebra of differential polynomials in $\omega(q_j)$, where $j\in J^f$ and  $\omega$ is the isomorphism \eqref{eq:omega}. In other words, $\widetilde{S}=\{\omega(q_j)|j\in J^f\}$ is a minimal strong generating set. On the other hand, $\widetilde{S}$ is not minimal as a weak generating set and we can find a proper weak generating set. 
 To verify that a given subset $S$ of $\widetilde{S}$ is a weak generating set, it is enough to show that for any $q_j$, there is a homogeneous element whose linear term is of the form $\omega(q_j)$ weakly generated by $S$.

\begin{ex}\label{ex:principal-1}
Let $\g=\mathfrak{sl}_{N}$ with $N\geq 3$, $f$ be a principal nilpotent element of $\g$ and $\tilde{k}\neq 0$.
Take the basis $\{q_{i}=\sum_{j=1}^{N+1-i}e_{j+i-1, j}|\, i=2, 3, \cdots, N\}$ of $\g^f$, where the conformal weight of ${q_i}$ is $i$. There are two different minimal weak generators of $\mathcal{W}^k(\g,f)$: 
\begin{enumerate}[]
\item \textup{(1) }$\{ \omega\big(q_3\big) \}$ : Observe that the element $\omega\big(q_{3}\big)_{(3)}\omega\big(q_3\big)$ is a  constant multiple of $\omega\big(q_2\big)$. 
Also, by Theorem \ref{thm:main lemma}, the linear term of $\omega\big(q_3\big)_{(1)}\omega\big(q_{i-1}\big)$ is a constant multiple of $\omega\big(q_{i}\big)$ for $3\leq i\leq N$.
\item \textup{(2-1) } $\{\omega\big(q_{N}\big)\}$ if $N$ is odd : 
For even $m\leq N$, consider $\omega\big(q_{N}\big)_{(2N-m-1)}\omega\big(q_{N}\big)$ to get an element whose linear term is $\omega\big(q_{m}\big)$ and consider $\big(\omega\big(q_{N}\big)_{(N)}\omega\big(q_{N}\big)\big)_{(2N-m-3)}\omega\big(q_{N}\big)$ to get an element whose linear term is $\omega\big(q_{m+1}\big)$ and the element $\omega\big(q_{m+1}\big)$ occurs as the linear term of the product $\big(\omega\big(q_{N}\big)_{(N)}\omega\big(q_{N}\big)\big)_{(2N-m-3)}\omega\big(q_{N}\big)$. 
\item \textup{(2-2) }$\{\omega\big(q_{N-1}\big)\}$ if $N$ is even : 
For even $m\leq N$, consider $\omega\big(q_{N-1}\big)_{(2N-m-3)}\omega\big(q_{N-1}\big)$ to get an element with $\omega\big(q_{m}\big)$ and consider $\big(\omega\big(q_{N-1}\big)_{(N-3)}\omega\big(q_{N-1}\big)\big)_{(2N-m-3)}\omega\big(q_{N-1}\big)$ to get an element with $\omega\big(q_{m+1}\big)$. 
\end{enumerate}
\end{ex}

\subsection{Quantum and classical W-algebras} \label{subsec:Q-C W-algebra}

In this subsection, we briefly review the definition of quantum W-algebras and their relationship to classical W-algebras. 
For detailed explanations, we refer to \cite{DSK06}. The notation used here is consistent with that in Section \ref{subsec:classical W}.

\begin{defn} \label{def:vA}
    A \textit{vertex algebra} is a quintuple $(V, \vac, \partial, [\ _\lambda \ ], :\ :)$ which satisfies the following axioms:
    \begin{enumerate}[(i)]
    \item $(V, \partial, [\ _\lambda \ ])$ is a LCA,
    \item $(V, \vac, \partial, :\ :)$ is a unital differential algebra with a derivation $\partial$, satisfying the following conditions:
    \begin{enumerate}[]
\item  \textup{(Quasi-commutativity)}
   $:ab:-(-1)^{p(a)p(b)}:ba:=\int_{-\partial}^{0} [a\ _\lambda\ b]d\lambda$,
\item \textup{(Quasi-associativity)}
     $:a:bc::-(-1)^{p(a)p(b)}:b:ac::=:\left(\int_{-\partial}^{0} [a\ _\lambda\ b]d\lambda\right) c:$,   
      \end{enumerate}
      for $a, b, c\in V$.
       Here, the operation $:\ :$ is called the \textit{normally ordered product} of $V$. 
    \item The $\lambda$-bracket and the normally ordered product are related by 
    \begin{enumerate}[]
    \item \textup{(Wick Formula)} $[a\, {}_\lambda\, :bc:]=:[a \, {}_\lambda\, b]c:+(-1)^{p(a)p(b)}:b[a\, {}_\lambda\, c]:+\int_{0}^{\lambda}[[a\, {}_\lambda\, b]\, {}_\mu\, c]d\mu$ for  $a,b,c\in V$.
    \end{enumerate}
    \end{enumerate}
\end{defn}

A subset $S$ of a vertex algebra $V$ is said to {\it strongly generate} $V$ if every element in $V$ can be obtained from $S$ by the normally ordered product and the derivation $\partial$. 
Similarly, a subset $S$ is said to {\it weakly generate} $V$ 
if every element in $V$ can be written as a linear combination of elements of the form $a^i_{(n_1)}\cdots a^s_{(n_s)}\vac$ where $a^i\in S$\,,\,$n_i\in \mathbb{Z}$ and $s\geq0$.
These notions will be used in the following section to analyze the structure of W-algebras and to describe their generating sets for various types of nilpotent elements. 

As in Poisson vertex algebras cases, whenever a LCA $R$ is given, one can construct a vertex algebra $V(R)$ called universal enveloping vertex algebras. Indeed, this result can be generalized by considering so-called nonlinear LCA $R$, which is a $\mathbb{C}[\partial]$-module with a $\lambda$-bracket whose image may not be in $R[\lambda]$ (see \cite{DSK06} for details).
To endow such a vertex algebra with a conformal grading, we define a conformal vector of the vertex algebra $V$.
An element $L\in V$ is called a {\it conformal vector} if it satisfies (i) $
[L\ _\lambda\ L]=(\partial+2\lambda)L+\frac{c}{12}
\lambda^3$
for some $c\in\CC$, (ii) $L_{(0)}=\partial$, (iii) $L_{(1)}$ acts semisimply on $V$. 
In this setting, the eigenvalue of the homogeneous element $a\in V$ under $L_{(1)}$ is called {\it conformal weight} of $a$.

\vskip 2mm

In order to define quantum affine W-algebras, we need the following nonlinear LCAs. 
\begin{enumerate}[(i)]
\item The nonlinear current LCA is strongly generated by $\{a\mid a\in\g\}$ with the $\lambda$-bracket \eqref{bracket of v^k(g)}.
\item Let $\phi_{\mathfrak{n}}$ be a vector superspace isomorphic to $\Pi(\mathfrak{n})$, where $\Pi$ denotes a parity reversing map and let $\phi^{\n^*}\simeq\Pi(\n^*)$.
Then the charged free fermion nonlinear LCA is strongly generated by $\{\phi_n, \phi^\theta\mid n\in\mathfrak{n}, \theta\in\n^*\}$ satisfying $[\phi_n\ _\lambda \ \phi^\theta]=\theta(n),\ [\phi_{n_1}\ _\lambda \ \phi_{n_2}]=0=[\phi^{\theta_1}\  _\lambda\ \phi^{\theta_2}]$ for $n_1, n_2\in\n$ and $\theta_1, \theta_2\in\n^*$. 
\item Let $\Phi_{\g_{1/2}}$ be a vector superspace isomorphic to $\g_{1/2}$. 
Then the neutral free fermion nonlinear LCA is strongly generated by $\{\Phi_c\mid c\in\g_{1/2}\}$ with the $\lambda$-bracket $[\Phi_c \ _\lambda \ \Phi_d]=(f\mid [c, d])$ for $c, d\in\g_{1/2}$.
\end{enumerate}
The universal enveloping vertex algebras of these nonlinear LCAs are denoted respectively 
by $V^k(\g)$, $\mathcal{F}^{ch}$ and $\mathcal{F}^{ne}$.
We endow the vertex algebra $C(\g,f,k):=V^k(\g)\otimes \mathcal{F}^{ch}\otimes \mathcal{F}^{ne}$ with a $\ZZ$-grading, referred to as the {\it charge}, by giving degree $1$ to $\phi
^\theta$, the degree $-1$ to $\phi_n$ and the degree $0$ to the other basis elements. 
It is known that $C(\g,f, k)$ has a conformal vector $L$ when $k \neq -h^\vee$ which is induced from the conformal vector $L_{\text{sug}}+\partial x$ of $V^k(\g)$, where $L_{\text{sug}}$ is the sugawara operator, together with the conformal vectors of $\mathcal{F}^{ch}$ and $\mathcal{F}^{ne}$. 
With respect to $L$, each element $a\in\g_{i}\subset V^k(\g)$ has the conformal weight $1-i$, element $\phi_n\in\mathcal{F}^{ch}$ has the  conformal weight $1-j$, where $n\in\g_{j}$, and element $\Phi_c\in\mathcal{F}^{ne}$ has the conformal weight $\frac{1}{2}$. 
In addition, for each $\theta\in\mathfrak{n}^*$, there exists an element $n_\theta\in\g_{l}$ such that $\theta=(n_\theta\mid\cdot)$ and then element $\phi^\theta\in\mathcal{F}^{ch}$ has the conformal weight $l$.

Let $\{u_\al\mid \al\in S_i\}$ be a basis of $\g_i$ consisting of elements homogeneous with respect to the parity and let $\{u^\al\}$ denote its dual basis with respect to $(\cdot\mid\cdot)$.
We denote by $S_{>0}=\mathop{\cup}_{i>0}{S_i}$ and $\phi^\al=\phi^{(u^\al\mid\cdot)}$ for $\al\in S_{>0}$, where $(u^\al\mid\cdot)\in\mathfrak{n}^*$.
In $C(\g, f, k)$, we define the following derivation
\begin{equation}\label{element d}
    \begin{aligned}
d=& \displaystyle\sum_{\al\in S_{>0}}(-1)^{p(u_\al)}:\phi^\al u_\al:+\displaystyle\sum_{\al\in S_{1/2}}:\phi^\al \Phi_{u_\al}:\\
& +\displaystyle\sum\limits_{\al\in S_{>0}}(f\mid u_\al)\phi^\al+\frac{1}{2}\sum\limits_{\al, \be\in S_{>0}}(-1)^{p(u_\al)}:\phi^\al\phi^\be\phi_{[u_\be, u_\al]}:,
\end{aligned}
\end{equation}
 of charge $1$ such that $d_{(0)}^2=0$.
Then $(C(\g, f, k), d_{(0)})$ is called the BRST complex.

\begin{defn}
The (quantum affine) W-algebra of level $k\in\CC$ associated with $\g$ and $f$ is 
\begin{align*}
W^k(\g, f)=H(C(\g, f, k), d_{(0)}).
\end{align*}
\end{defn}
\nin Note that the W-algebra $W^k(\g, f)$ admits a conformal vector $L$, induced from the conformal vector $L$ of $C(\g, f, k)$, provided $k\neq-h^{\vee}$. 

Kac-Wakimoto \cite{KW} showed that $H(C(\g, f, k), d_{(0)})$ is concentrated in charge $0$. 
More precisely, for each element $a\in\g$, they introduced a building block $J_a:=a+\sum_{\al\in S_{>0}}:\phi^\al\phi_{[u_\al, a]}:$, which has the conformal weight $1-i$ where $a\in\g_i$. Then $C(\g, f, k)$ can be viewed as the vertex algebra generated by  $J_a, \phi_{u_\al}, \phi^\al$ and $\Phi_{u_\be}$ for $a\in\g$, $\al\in S_{>0}$ and $\be\in S_{1/2}$.
Using the Künneth lemma, $H(C(\g, f, k), d_{(0)})$ is isomorphic to the tensor product of two cohomologies $H(C^{-}, d_{(0)}|_{C^{-}})$ and $H(C^{+}, d_{(0)}|_{C^{+}})$, where $C^{-}\subset C(\g, f, k)$ is generated by $J_a$, $\phi^\al$ and $\Phi_{u_\be}$ for $a\in\g_{\leq0}=\oplus_{i\leq 0}\g_{i}, \,\al\in S_{>0}$ and $\be\in S_{1/2}$ and $C^{+}\subset C(\g, f, k)$ is generated by $\phi_{u_\al}$ and $d_{(0)}\phi_{u_\al}$ for $\al\in S_{>0}$.  The latter cohomology is isomorphic to $\CC$ and only the former cohomology remains. Finally, it can be shown that $H(C^{-}, d_{(0)}|_{C^{-}})=H^0(C^{-}, d_{(0)}|_{C^{-}})$. Moreover, the following theorem holds.

\begin{thm}[\cite{KW}]\label{explicit form of generator of affine}
Let $a$ be an element in $\g^f$ with conformal weight $\Delta_a.$
\begin{enumerate}[(1)]
\item There exists a weight $\Delta_a$ element $J_{\{a\}}$ in $W^k(\g,f)\subset C^-$ such that $J_{\{a\}}-J_a$ is in the differential algebra generated by $J_b$ and $\Phi_{u_\be}$ for $b\in\g_{\leq0}$ and $\be\in S_{1/2}$, whose conformal weights are strictly lower than $\Delta_a$. 
\item Let $\{a_i\}$ be a homogeneous basis of $\g^f$. 
Then $\{J_{\{a_i\}}\}$ strongly generates the W-algebra $W^k(\g, f)$.
\end{enumerate}
\end{thm}

In order to describe the relation between quantum affine and classical affine W-algebras, we briefly recall the notion of the quasi-classical limit of a vertex algebra. 
Let $(V_{\epsilon}, \vac_{\epsilon}, \partial_\epsilon, [\cdot \ _\lambda \ \cdot]_\epsilon, :\ :_\epsilon)$ be a family of vertex algebras over $\CC[\epsilon]$ where $V_\epsilon$ is free as a $\CC[\epsilon]$-module and $[V_\epsilon \ _\lambda \ V_\epsilon]\subset\CC[\lambda]\otimes\epsilon V_\epsilon$.
Then the quotient $V^{\mathrm{cl}}:=V_{\epsilon}/\epsilon V_\epsilon$ is a supercommutative differential algebra. 
For lifts $\bar{a}, \bar{b}$ of $a, b\in V^{\mathrm{cl}}$, we define 
\begin{equation}\label{eq:bracket_classical}
\{a\, {}_\lambda\, b\}:=\frac{1}{\epsilon}[\bar{a}\, {}_\lambda \ \bar{b}]_\epsilon |_{\epsilon=0}, \quad ab \, := \, :\bar{a}\bar{b}:_{\epsilon}.
\end{equation}
It is known by  \cite{DSK06} that this operation endows the quotient with the structure of a Poisson vertex algebra. 
In this case, we say that we take the \textit{quasi-classical limit} of the vertex algebra $V$.

In some articles such as \cite{FB01,CL22}, it is shown that the classical W-algebra $\mathcal{W}^k(\g,f)|_{k=0}$ can be constructed using a classical limit of BRST complex. 
However, in order to get the classical W-algebra $\mathcal{W}^k(\g,f)$ for nonzero $k\in \mathbb{C}$ using BRST complex, we need a slightly different complex 
\begin{equation} \label{eq:complex-with parameter}
    C(\g,f,k)_{\epsilon}=\mathbb{C}[\epsilon]\otimes C(\g,f,k),
\end{equation} 
which is introduced in \cite{DSK06}. The OPE of $C(\g,f,k)_{\epsilon}$ is defined as follows: 
\begin{equation} \label{eq:bracket_epsilon_second}
      [a{}_\lambda b]_{\epsilon}= \epsilon([a,b])+ \epsilon k(a|b)\lambda, \quad [\phi_n{}_\lambda \phi^\theta]_{\epsilon}=\epsilon\theta(n),  \quad [\Phi_{m_1}{}_\lambda \Phi_{m_2}]_\epsilon=\epsilon (f|[m_1,m_2]).
\end{equation}
Consider 
\begin{equation}\label{element d-epsilon}
    \begin{aligned}
d^{\epsilon}_k=& \displaystyle\sum_{\al\in S_{>0}}(-1)^{p(u_\al)}:\phi^\al u_\al:+\displaystyle\sum_{\al\in S_{1/2}}:\phi^\al \Phi_{u_\al}:\\
& +\displaystyle\sum\limits_{\al\in S_{>0}}(f\mid u_\al)\phi^\al+\frac{1}{2}\sum\limits_{\al, \be\in S_{>0}}(-1)^{p(u_\al)}:\phi^\al \phi^\be \phi_{[u_\be, u_\al]}:\, \in C(\g,f,k)_{\epsilon}.
\end{aligned}
\end{equation}
Then $Q^\epsilon:= \frac{1}{\epsilon}d^\epsilon_k{}_{(0)}$ is a differential on $C(\g,f,k)_{\epsilon}$.
It is obvious that 
$ H(C(\g,f,k)_{\epsilon}, Q^{\epsilon})\big|_{\epsilon=1} \simeq W^k(\g,f).$
Now the classical limit of the vertex algebra $C(\g,f,k)_{\epsilon}$ is the classical BRST complex 
\begin{equation} \label{eq:classical_complex}
    \mathcal{C}(\g,f,k)\simeq S\Big(\mathbb{C}[\partial]\otimes\big(\g\oplus\phi_{\n}\oplus \phi^{\n^*} \oplus \Phi_{\g_{1/2}}\big)\Big)
\end{equation}
and the Poisson $\lambda$-bracket on $\mathcal{C}(\g,f,K)$ is defined by 
\begin{equation}
    \{a{}_\lambda b\}= [a,b]+K(a|b)\lambda, \quad \{\phi_n {}_\lambda \phi^\theta\}=\theta(n), \quad \{\Phi_{m_1}{}_\lambda \Phi_{m_2}\}=(f|[m_1,m_2]).
\end{equation}
 Finally, for the element $d^{cl}_k \in \mathcal{C}(\g,f,k)$ which is induced from the element $d^\epsilon_k$, its $0$-mode $\mathcal{Q}:=d^{cl}_{k\, (0)}$ is a differential and 
$    H(\mathcal{C}(\g,f,k), \mathcal{Q}) = H(C(\g,f,k)_{\epsilon}, Q^{\epsilon})\big|_{\epsilon=0} \simeq \mathcal{W}^k(\g,f).$
For more detailed explanations for this paragraph, we refer to Section 6 of \cite{DSK06}. Later, we will use the following theorem to describe weak generators of quantum W-algebras.

\begin{thm}\cite{DSK06} \label{thm:quantum-classical}
    Let $C(\g,f,k)_\epsilon$ and $Q^{\epsilon}$ be the complex and differential defined as above. Then the cohomology $W^k(\g,f)_\epsilon:= H(C(\g,f,k)_\epsilon,Q^\epsilon)$ is a vertex algebra over $\mathbb{C}[\epsilon].$ Moreover,  
    \[W^k(\g,f)_\epsilon|_{\epsilon=1}=W^k(\g,f), \quad W^k(\g,f)_\epsilon|_{\epsilon=0}=\mathcal{W}^k(\g,f),\]
    as vertex algebras and Poisson vertex algebras, respectively.
\end{thm}

\section{Weak Generating Set of W-algebras of type $A$}\label{Chap:large}
When the nilpotent element $f$ has a single Jordan block, Example \ref{ex:principal-1} shows the classical affine W-algebra $\mathcal{W}^k(\g,f)$ admits two different weak generating sets, each consisting of a single element.
This simple case serves as a model for understanding how the structure of $\mathcal{W}^k(\g,f)$ extends to more general types of $f$. 

Let $\g=\mathfrak{sl}_N$ or $\mathfrak{sl}_{N_1|N_2}$, where $N=N_1+N_2$, and let $f$ be an even nilpotent whose Jordan block is of type $(m_1, m_2, \cdots, m_d)$.
In the purely even case $\g=\mathfrak{sl}_N$, the element $f$ satisfies 
\begin{equation}\label{eq:f}
N=m_1+m_2+\cdots+m_d\ , \ m_1\geq m_2\geq\cdots\geq m_d\ , \ d\geq2\ ,
\end{equation}
so that nilpotent orbits are parametrized by a single partition of $N$.
By contrast, for $\g=\mathfrak{sl}_{N_1|N_2}$, nilpotent orbits are classified by a pair of partitions$-$one for $N_1$ and the other for $N_2-$corresponding to the even and odd parts of the superalgebra. 

In this section, we find two distinct weak generating sets of $\mathcal{W}^k(\g,f)$ for $k \neq 0$. 
One is a set consisting of large conformal weight elements and the other one is a set with small conformal weight elements. Throughout this section, since all statements and proofs apply uniformly for all $k\neq 0,$ we fix $k$ such that 
\begin{equation}
    k(e|f)_{nor}=\frac{k}{2}(h|h)_{nor} =1.
\end{equation}
In other words, we assume the bracket on $\mathcal{W}^k(\g,f)$ is induced from $\{a{}_\lambda b\}= [a,b]+\lambda(a|b)$ for $a,b\in \g.$

\subsection{Basis of the centralizer of $f$}\label{Subsec:basis of gf}
In this subsection, we explicitly describe a basis of $\g^f$. Let us visualize an element of $\g$ as a block matrix. The $N\times N$ matrix can be visualized as divided into $d^2$ blocks, as 
\begin{equation} \label{eq:block matrix}
   \begin{bmatrix} 
   B^{(1,1)} & B^{(1,2)} & \cdots & B^{(1,d)}  \\
   B^{(2,1)} & B^{(2,2)} & \cdots &  B^{(2,d)}  \\
   \vdots & \vdots & \ddots &  \vdots  \\
   B^{(d,1)} & B^{(d,2)} &  \cdots &  B^{(d,d)} \\
   \end{bmatrix} ,
\end{equation}
where $B^{(i, j)}$ is an $m_i\times m_j$ matrix. We use the matrix \eqref{eq:block matrix} for both $\g=\mathfrak{sl}_N$ and $\g=\mathfrak{sl}_{N_1|N_2}$ cases.
Let 
\begin{align}
    e^{(i, j)}_{a, b}\in\mathfrak{gl}_N \text{ or } \mathfrak{gl}_{N_1|N_2} \ (1\leq i,j\leq d),
\end{align}
denote the matrix having a single nonzero entry $1$ at the $(a, b)$-position in the block $B^{(i, j)}$ and zeros elsewhere, including all other blocks. 
Then the nilpotent element $f$ of type \eqref{eq:f} is   
\begin{equation} \label{eq:f-1}
    f=\sum_{j=1}^{d}\sum\limits_{i=1}^{m_j-1}e_{i+1, i}^{(j, j)}\ \in \ \g_{-1}.
\end{equation}
The element $e$ in the $\mathfrak{sl}_2$-triple is 
\begin{equation} \label{eq:e}
e=\frac{1}{M}(\mathsf{e}^{(1,1)}+\mathsf{e}^{(2,2)}+ \cdots+ \mathsf{e}^{(d,d)}) ,   
\end{equation}
where 
\begin{equation} \label{eq:e,j}
    \mathsf{e}^{(j,j)}= \sum_{1\leq i \leq m_j-1} (i m_j-i^2)e^{(jj)}_{i, i+1} 
\end{equation}
and 
\begin{equation} \label{eq:M}
    M=\sum_{j=1}^d M_j  \quad \text{ for } \quad  M_j= \frac{(m_j-1)m_j (m_j+1)}{6}.
\end{equation}


Let $j,l$ be distinct integers such that $1\leq j, l\leq d$ and let $t_{jl}=t-\frac{\lvert m_j-m_l\rvert}{2}$ also be an integer satisfying $1\leq t_{jl}\leq \mathrm{min}\{m_j, m_l\}$.
Denote an element of $\g^f_{1-t}$ contained in the block $B^{(j, l)}$ by 
\begin{equation} \label{eq:(j,l)}
q^{ (j,l)}_{\ t}=\sum_{i=1}^{1+\frac{m_j+m_l}{2}-t}e_{t'_{jl}+(i-1),\,  i}^{\ \, ( j ,  l )} \, ,
 \end{equation}
 where $t'_{jl}=t+\frac{m_j-m_l}{2}.$
In diagonal blocks, there are more centralizers denoted by  
\begin{equation}\label{eq:(j,j)}
     q^{(j, j)}_{\ t}=\left\{\begin{array}{ll}\sum_{i=1}^{m_1}m_j e_{i, i}^{(1, 1)}-(-1)^{p(e^{(j, 1)}_{1,1})}\sum_{i=1}^{m_j}m_1 e_{i, i}^{(j, j)} & \text{ if } t=1.  \\
      \sum_{i=1}^{1+m_j-t}e_{t+(i-1), i}^{(j,j)}  & \text{ if } t=2, \cdots, m_j.\end{array} \right.
\end{equation}
In the following sections, we will use the basis of the centralizer in the following lemma.

\begin{lem}  \label{lem:basis g^f}
For the elements in $\g$ given in \eqref{eq:(j,j)} and \eqref{eq:(j,l)}, the set 
\begin{equation}
\begin{aligned}
     \mathcal{B}^f :=  \{\, q^{ (j,j)}_{\ 1} \, |\, j=2,\cdots, d\, \}  \, \cup \, &  \{\, q^{ (j,j)}_{\ t} \, |\, j=1,\cdots, d, \ t=2, \cdots, m_j \, \} \\
        & \cup  \{\, q^{ (j,l)}_{\ t}\,  |\, 1\leq j\neq l\leq d,  \ 1\leq t_{jl}\leq \text{min}\{m_j, m_l\} \}
\end{aligned}
\end{equation}
where $t_{jl}=t-\frac{|m_j-m_l|}{2}$ is a basis of $\g^f.$
\end{lem}
\begin{proof}
By direct computations, one can check each element in $\mathcal{B}^f$ is in $\g^f.$ 
Hence it is enough to check the number of elements in $\mathcal{B}^f$ is the same as $\text{dim}\g^f.$ 
More precisely, since $q_s^{(i,j)}\in \g^f_{1-s}$, we need to prove that  
$\text{dim}\big(\, \g^f\, \cap\ \g^{(i,j)}_{\frac{m_j-m_i}{2}-t}\, \big)=1$
for $t=0, \cdots, m_j-1$ when $m_j\leq m_i.$
In order to show that, let us denote by $\g^{(i,j)}_{s}$ the subset of $\g_s$ in $B^{(i,j)}.$ One can easily see that 
\[\text{dim}\, \g^{(i,j)}_{\frac{m_j-m_i}{2}-t}=m_j-t, \quad \text{dim}\, \g^{(i,j)}_{s}=m_j \]
for $t=0, \cdots, m_j-1$ and $\frac{m_j-m_i}{2}<s<0.$ Hence the dimension of $\text{dim}\big(\, \g^f\, \cap\ \g^{(i,j)}_{\frac{m_j-m_i}{2}-t}\, \big)=1$ for $t=0, \cdots, m_j-1.$ 
\end{proof}

For the isomorphism $\omega$ in \eqref{eq:omega}, the set $\omega(\mathcal{B}^f)$ strongly generates $\mathcal{W}^k(\g,f)$ by Lemma \ref{lem:basis g^f} and the conformal weight of each element is given by 
\begin{equation} \label{Eq:conf wt of gen}
    \Delta_{\omega\big(q_{\ t}^{(j,j)}\big)}=\Delta_{\omega \big( q_{\ t}^{(j,l)}\big)}=t. 
\end{equation}
The following notations are introduced to use Theorem \ref{thm:main lemma} in the subsequent sections.

\begin{notation} \label{notation}
    Recall $q^{(i,j)}_{1-t}\in \g_t$ and the conformal weight of $q^{(i,j)}_{1-t}$ is $1-t.$
    \begin{itemize}
    \item  The index set $J^f_t$ consists of the $(t,i,j)$ where $q^{(i,j)}_{1-t}$ is in $\mathcal{B}^f$ and $J^f=\cup_{t\in \mathbb{Z}/2} J^f_t.$ 
    \item For $(t,i,j)\in J^f$, we identify $q_{(1-t,i,j)}[0]$ in Theorem \ref{thm:main lemma} with $q^{(i,j)}_{1-t}$ in Lemma \ref{lem:basis g^f}.
    \item For $(t,i,j)\in J^f$, we consider the basis $\mathcal{B}^e$ of $\g^e$ consisting of the dual elements $q^{(i,j)*}_{\, 1-t}$ of $q^{(i,j)}_{1-t}\in \mathcal{B}^f$. 
    \item Let $q^{(i,j)*}_{\, 1-t}[n]:=(\text{ad} f)^n q^{(i,j)*}_{\, 1-t}$ 
    and $q^{(i,j)}_{\, 1-t}[n]$ be the dual of $q^{(i,j)*}_{\, 1-t}[n]$, so that the following two bases of $\g$ are dual with respect to $(\ | \ )$:
    \begin{equation}
        \begin{aligned}
             & \mathcal{B}^*=\{\, q^{(i,j)*}_{\, 1-t}[n] \, |\,  (i,j,t)\in J^f, n=0,1,\cdots,-2t\, \}, \\
             & \mathcal{B}=\{\, q^{(i,j)}_{\, 1-t}[n] \, |\,  (i,j,t)\in J^f, n=0,1,\cdots,-2t\, \}.
        \end{aligned}
    \end{equation}
\end{itemize}
\end{notation}

\subsection{Weak generating set of $\mathcal{W}^k(\g,f)$ with large conformal weight}\label{Subsec:large}
In this subsection, we describe a weak generating set consisting of large conformal weight elements. Assume $\g=\mathfrak{sl}_N$ and $f$ is \eqref{eq:f-1} and we use Notation \ref{notation}. The following theorem is the main result in this subsection.

\begin{thm}\label{big thm}
Recall that $f$ is the nilpotent with the partition $(m_1, m_2, \cdots, m_d)$ and $\omega$ is the isomorphism given in 
\eqref{eq:omega}.
\begin{enumerate}[(1)]
\item If $m_1>m_2$, then the following $2d-2$ elements
\begin{align}\label{weak generators}
\omega\Big(q_{\frac{m_{i}+m_{i+1}}{2}}^{(i,i+1)}\Big)\,,\,  \omega\Big(q_{\frac{m_{i}+m_{i+1}}{2}}^{(i+1, i)}\Big),\,  (1\leq i\leq d-1),
\end{align}
weakly generate $\mathcal{W}^k(\g,f)$ for $k\neq0$.
\item If $m_1=m_2$, then the following $2d-2$ elements
\begin{align}\label{weak generators m_1=m_2}
\omega\Big(q_{\,{m_1}-1}^{(2, 1)}\Big)\,,\,  \omega\Big(q_{\frac{m_{i}+m_{i+1}}{2}}^{(i,i+1)}\Big)\,,\,   \omega\Big(q_{\frac{m_{j}+m_{j+1}}{2}}^{(j+1, j)}\Big),\,  (1\leq i\leq d-1, \ 2\leq j\leq d-1),
\end{align}
weakly generate $\mathcal{W}^k(\g,f)$ for $k\neq0$.
\end{enumerate}
\end{thm}

Recall that $\mathcal{W}^k(\g,f)$ is the polynomial of differential algebras in $\omega(\mathcal{B}^f).$
As we mentioned in Section \ref{sec:prelim}, any subset of $\mathcal{W}^k(\g,f)$ consisting of homogenous elements whose linear part without total derivatives spans $\omega(\g^f)$ can be a strong generating set. Hence it is enough to check if the given elements weakly generate a homogeneous element having $\omega\big( q^{(i,j)}_{1-t}\big)$ for any $(t,i,j)\in J^f.$
In Lemma \ref{lem:Thm-1}--\ref{lem:Thm-4}, detailed proof will be provided. Before that, we briefly explain how to get all the strong generators from the elements in Theorem \ref{big thm}. 

\vskip 3mm 

\noindent \textbf{Outline of the Proof}   
\begin{enumerate}[\ (\text{Step}1)]
\item  Get an element  having $c_1\omega\big(q_{\,2}^{(1, 1)}\big)+c_2\omega\big(q_{\,2}^{(2, 2)}\big)$ for $c_1, c_2\in \mathbb{C}$:
\begin{itemize}
\item If $m_1>m_2$, get this element from  $\omega\big(q_{\frac{m_1+m_2}{2}}^{(1,2)}\big)_{(m_1+m_2-3)}\omega\big(q_{\frac{m_1+m_2}{2}}^{(2,1)}\big)$. 
\item If $m_1=m_2$, get this element  from $\omega\big(q_{\, m_1}^{(1,2)}\big)_{(2m_1-4)}\omega\big(q_{\, {m_1}-1}^{(2,1)}\big)$.
\end{itemize}
\item  Get $\omega\big(q_{\frac{m_1+m_2}{2}-1}^{(1,2)}\big)$ and $\omega\big(q_{\frac{m_1+m_2}{2}-1}^{(2,1)}\big)$:
\begin{itemize}
\item Consider $\big(c_1\omega\big(q_{\,2}^{(1, 1)}\big)+c_2\omega\big(q_{\,2}^{(2, 2)}\big)\big)_{(2)}\omega\big(q_{\frac{m_1+m_2}{2}}^{(1,2)}\big)$ to get an element $\omega\big(q_{\frac{m_1+m_2}{2}-1}^{(1,2)}\big)$.
\item  If $m_1>m_2$, consider $\big(c_1\omega\big(q_{\,2}^{(1, 1)}\big)+c_2\omega\big(q_{\,2}^{(2, 2)}\big)\big)_{(2)}\omega\big(q_{\frac{m_1+m_2}{2}}^{(2,1)}\big)$ to get $\omega\big(q_{\frac{m_1+m_2}{2}-1}^{(2,1)}\big)$. 
\item If $m_1=m_2$, we already have $\omega\big(q_{\frac{m_1+m_2}{2}-1}^{(2,1)}\big)$.
\end{itemize} 



\item  Get an element  having $\omega\big(q_{\,2}^{(j,j)}\big)$ for $1\leq j\leq d$ : 
\begin{itemize}
\item
Consider $\omega\big(q_{\frac{m_1+m_2}{2}-1}^{(1, 2)}\big)_{(m_1+m_2-5)}\omega\big(q_{\frac{m_1+m_2}{2}-1}^{(2,1)}\big)$ to get  $d_1\omega\big(q_{\,2}^{(1, 1)}\big)+d_2\omega\big(q_{\,2}^{(2, 2)}\big)$ for $d_1, d_2\in \mathbb{C}$.
\item Combine with the element in (Step1).
\item All other generators can be obtained in a similar way. 
\end{itemize}
\item  Get an element  having $\omega\big(q_{\,t}^{(1, 2)}\big)$ or $\omega\big(q_{\,t}^{(2, 1)}\big)$ for $ t_{12}\in\{1,\cdots,m_2-2,m_2\}$, where $t_{12}=t-\frac{m_1-m_2}{2}$ :
\begin{itemize}
\item If $2\leq t_{12}\leq m_2-1$, 
consider $\omega\big(q_{\,2}^{(1,1)}\big)_{(2)}\omega\big(q_{\,t}^{(1,2)}\big)$ and $\omega\big(q_{\,2}^{(1,1)}\big)_{(2)}\omega\big(q_{\,t}^{(2,1)}\big)$ to get elements $\omega\big(q_{\,t-1}^{(1,2)}\big)$ and $\omega\big(q_{\,t-1}^{(2,1)}\big)$.
\item If $m_1>m_2$, we already have $\omega\big(q_{\frac{m_1+m_2}{2}}^{(2, 1)}\big)$.
\item If $m_1=m_2$, consider $\omega\big(q_{\,2}^{(1,1)}\big)_{(0)}\omega\big(q_{\,m_1-1}^{(2, 1)}\big)$ to get an element  with $\omega\big(q_{\,m_1}^{(2, 1)}\big)$. 
\end{itemize}

\item  Get an element  having $\omega\big(q_{\,t}^{(j, l)}\big)$ for $j\neq l$ with $(j, l)\neq(1, 2), (2, 1)$ and $1\leq t-\lvert\frac{m_j-m_l}{2}\rvert\leq \text{min}\{m_j, m_l\}$:
\begin{itemize}
\item
Consider $\big(\big(\omega\big(q_{\,2}^{(1,1)}\big)_{(2)}\big)^{m_2-2}\omega\big(q_{\frac{m_1+m_2}{2}-1}^{(2,1)}\big)\big)_{(0)}\big(\big(\omega\big(q_{\,2}^{(2,2)}\big)_{(2)}\big)^{m_3-1}\omega\big(q_{\frac{m_2+m_3}{2}}^{(3,2)}\big)\big)$ to get an element with $\omega\big(q_{\frac{m_1-m_3}{2}+1}^{(3,1)}\big)$.
\item If $2\leq t\leq m_3$, consider $\big(\omega\big( q_{\, 2}^{(1,1)}\big)_{(0)}\big)^{t-1}\omega\big(q_{\frac{m_1-m_3}{2}+1}^{(3,1)}\big)$ to get an element with
$\omega\big(q_{\frac{m_1-m_3}{2}+t}^{(3,1)}\big)$.

\item All other generators can be obtained in a similar way.
\end{itemize} 
\item Get an element having $\omega\big(q_{\,t}^{(j, j)}\big)$ for $t=1$ or $3\leq t\leq m_j$ :
\begin{itemize}
\item If $m_1>m_2$ and either $t=1$ or $3\leq t\leq m_2$, consider $\omega\big(q_{\frac{m_1+m_2}{2}}^{(1, 2)}\big)_{(m_1+m_2-t-1)}\omega\big(q_{\frac{m_1+m_2}{2}}^{(2, 1)}\big)$ and $\omega\big(q_{\frac{m_1+m_2}{2}-1}^{(1, 2)}\big)_{(m_1+m_2-t-3)}\omega\big(q_{\frac{m_1+m_2}{2}-1}^{(2, 1)}\big)$ to get an element with $\omega\big(q_{\,t}^{(2, 2)}\big)$.
\item If $m_1=m_2$ and either $t=1$ or $3\leq t\leq m_2$, consider $\omega\big(q_{\,m_1}^{(1, 2)}\big)_{(2m_1-t-2)}\omega\big(q_{\,m_1-1}^{(2, 1)}\big)$ and $\omega\big(q_{\,m_1-1}^{(1, 2)}\big)_{(2m_1-t-3)}\omega\big(q_{\,m_1-1}^{(2, 1)}\big)$ to get an element with $\omega\big(q_{\,t}^{(2, 2)}\big)$.
\item All other generators can be obtained in a similar way.
\end{itemize} 
\end{enumerate}
Through (Step1) to (Step6), we get a set of strong generators of $\mathcal{W}^k(\g,f)$ and hence the theorem is proved. \qed

\vskip 3mm 
Now, we show the detailed proof of each step in Lemma \ref{lem:Thm-1}--\ref{lem:Thm-4}, using Theorem \ref{thm:main lemma}. Throughout this paper, we use the symbol $\sim$ to denote equality up to a nontrivial constant multiple. 
Namely, for two elements $A, B$, we write 
\begin{equation} \label{eq:sim}
    A\sim B \quad \text{ if and only if } \quad  A=cB 
\end{equation}
for some $c\in\CC^\times$. 

\begin{lem}[Step 1]\label{lem:Thm-1}
Let us view the classical W-algebra  $\mathcal{W}^k(\g,f)$ as a polynomial algebra in $\bigsqcup_{i\in \mathbb{Z}_+}\partial^i \omega(\mathcal{B}^f)$, where $\mathcal{B}^f$ is in Lemma \ref{lem:basis g^f}.
\begin{enumerate}[(1)]
\item Suppose $m_1>m_2$. The linear term of  $\omega\big(q_{\frac{m_1+m_2}{2}}^{(1, 2)}\big)_{(m_1+m_2-3)} \omega\big(q_{\frac{m_1+m_2}{2}}^{(2,1)}\big)$ is $c_1\omega\big(q_{\,2}^{(1, 1)}\big)+c_2\omega\big(q_{\,2}^{(2, 2)}\big)$ for some $c_1, c_2\in \mathbb{C},$ up to a total derivative part.
\item Suppose $m_1=m_2$. The linear term of $\omega\big(q_{\,m_1}^{(1, 2)}\big)_{(2m_1-4)}\omega\big(q_{\,m_1-1}^{(2,1)}\big)$ is $c_1\omega\big(q_{\,2}^{(1, 1)}\big)+c_2\omega\big(q_{\,2}^{(2, 2)}\big)$ for some $c_1, c_2\in \mathbb{C}$,  up to a total derivative part.
\end{enumerate}
\end{lem}

\begin{proof}
(1)
In order to see the terms in $\omega\big(q_{\frac{m_1+m_2}{2}}^{(1, 2)}\big)_{(m_1+m_2-3)} \omega\big(q_{\frac{m_1+m_2}{2}}^{(2,1)}\big)$, let us use Theorem \ref{thm:main lemma}. Observe that the formula in Theorem \ref{thm:main lemma} can be simply rewritten as 
\begin{equation}\label{conformal weight m_1+m_2/2 bracket} 
\begin{aligned}
    & \Big\{\, \omega\Big(q_{\frac{m_1+m_2}{2}}^{(1,2)}\Big) \ {}_\lambda \ \omega \Big(q_{\frac{m_1+m_2}{2}}^{(2,1)}\Big)\, \Big\}\\
    & =
    -\sum_{p= -1}^{N}\prod_{\substack{{t=0,\cdots, p+1},\\(j_{t-1}, n_{t-1})\prec(j_{t}, n_{t})}} \Big(\omega \big(\, \big[\, q_{j_{t-1}}[{n_{t-1}+1}]\, ,\,  q_{j_t}^*[n_t]\, \big]^\sharp\, \big)-( \, q_{j_{t-1}}[{n_{t-1}+1}]\, \big| \,  q_{j_t}^*[n_t]\, )\, (\lambda+\partial)\Big),
\end{aligned}
\end{equation}
for $N>\!>0$.   In our case, $q^{(1,2)}_{\frac{m_1+m_2}{2}}=q_{j_{p+1}}^*[n_{p+1}]$ and $q_{\frac{m_1+m_2}{2}}^{(2,1)}=q_{j_{-1}}[{n_{-1}+1}]$ and the RHS of \eqref{conformal weight m_1+m_2/2 bracket} for $p=-1$ is $\omega\big( \big[\big(q_{\frac{m_1+m_2}{2}}^{(2,1)}\big), \big(q_{\frac{m_1+m_2}{2}}^{(1,2)}\big) \big]\big).$ 
Notice that (i) $q_{\frac{m_1+m_2}{2}}^{(1,2)}\sim q_{\frac{m_1+m_2}{2}}^{(2,1)*}[m_1+m_2-2]$
(ii) a nontrivial multiple of $\lambda$ appears when 
$( \, q_{j_{t-1}}[{n_{t-1}+1}]\, \big| \,  q_{j_t}^*[n_t]\, )\neq 0$, that is $j_{t-1}=j_t$ and $n_{t-1}+1=n_t$ (iii) the linear part without the derivation $\partial$ in the coefficient $\lambda^{m_1+m_2-3}$ can appear only when $p=m_1+m_2-4$, so that \eqref{conformal weight m_1+m_2/2 bracket}  is a product of $m_1+m_2-2$ terms. 
As a conclusion, the terms we need for this lemma in \eqref{conformal weight m_1+m_2/2 bracket} is 
\begin{equation} \label{eq:lem:Thm-1--1}
    -\sum_{n=0}^{m_1+m_2-3}(-\lambda)^{m_1+m_2-3}\omega \big(\, \big[\, q_{\frac{m_1+m_2}{2}}^{(2, 1)}[n]\, ,\,  q_{\frac{m_1+m_2}{2}}^{(2, 1)*}[n+1]\, \big]^\sharp\, \big).
\end{equation}
To simplify \eqref{eq:lem:Thm-1--1}, we use 
 \begin{align*}
\Big[\, {q}_{\frac{m_1+m_2}{2}}^{(2,1)}[n]\, ,\,  q_{\frac{m_1+m_2}{2}}^{(2,1)*}[n+1]\, \Big]^\sharp
=\frac{(n+1)(m_1+m_2-2-n)}{m_1+m_2-2}  \Big[{q}_{\frac{m_1+m_2}{2}}^{(2,1)}, q_{\frac{m_1+m_2}{2}}^{(2,1)*}[1]\Big]^\sharp,
\end{align*} 
and 
thus the linear term of $\omega\big(q_{\frac{m_1+m_2}{2}}^{(1,2)}\big)_{(m_1+m_2-3)}\omega\big(q_{\frac{m_1+m_2}{2}}^{(2,1)}\big)$ without $\partial$ is 
\begin{equation*}
    \begin{aligned}
        &  \sum\limits_{n=0}^{m_1+m_2-3}
        \omega \Big(\Big[{q}_{\frac{m_1+m_2}{2}}^{(2,1)}[n]\, ,\,  q_{\frac{m_1+m_2}{2}}^{(2,1)*}[n+1]\Big]^\sharp\Big)\sim\omega \Big(\Big[\, {q}_{\frac{m_1+m_2}{2}}^{(2,1)}, q_{\frac{m_1+m_2}{2}}^{(2,1)*}[1]\, \Big]^\sharp\Big).
    \end{aligned}
\end{equation*}
Recall that 
${q}_{\frac{m_1+m_2}{2}}^{(2,1)}=e_{m_2, 1}^{(2, 1)}$ and $q_{\frac{m_1+m_2}{2}}^{(2,1)*}[1]\sim(e_{1, m_2-1}^{(1, 2)}-e_{2,m_2}^{(1, 2)})$.
Hence
 $\big[{q}_{\frac{m_1+m_2}{2}}^{(2,1)}, q_{\frac{m_1+m_2}{2}}^{(2,1)*}[1]\big]^\sharp\sim \big(e_{2,1}^{(1, 1)}+e_{m_2, m_2-1}^{(2, 2)}\big)^\sharp\sim\big(\frac{1}{m_1(m_1+1)}q_{\,2}^{(1, 1)}+\frac{1}{m_2(m_2+1)}q_{\,2}^{(2, 2)}\big).$ Here we used the fact that $\mathsf{e}^{(j,j)}/M_j$ for $\mathsf{e}^{(j,j)}$ and $M_j$ in \eqref{eq:e,j} and \eqref{eq:M}  is the dual element of $q_{\,2}^{(j, j)}$. Additionally, we used $(e_{2,1}^{(1, 1)}|\mathsf{e}^{(1,1)}/M_1)= \frac{6}{m_1(m_1+1)}$ and $(e_{m_2, m_2-1}^{(2, 2)}|\mathsf{e}^{(2,2)}/M_2)=\frac{6}{m_2(m_2+1)}$.
Finally, we conclude 
\begin{align*}
\eqref{eq:lem:Thm-1--1}\sim\lambda^{m_1+m_2-3}\Big(\, \omega\big(q_{\,2}^{(1, 1)}\big)+\frac{m_1(m_1+1)}{m_2(m_2+1)}\omega\big(q_{\,2}^{(2, 2)}\big)\, \Big),
\end{align*}
and hence we get the lemma.
\\
\nin(2)
As in the proof of the first statement, if we denote by $A$ the linear term of $\omega(q_{\,m_1}^{(1,2)}) _{(2m_1-4)} \omega(q_{\,m_1-1}^{(2,1)})$ without $\partial$, then
\begin{align*}  
A\sim\omega\Big(\,\big[{q}_{\,m_1-1}^{(2,1)},\, q_{\,m_1}^{(2,1)*}[2]\big]^\sharp\,\Big).
\end{align*}
Recall that ${q}_{\,m_1-1}^{(2,1)}=e_{m_1-1, 1}^{(2, 1)}+e_{m_1,2}^{(2, 1)}$ and $q_{\,m_1}^{(2,1)*}[2]\sim\big(e_{3, m_1}^{(1, 2)}-2e_{2, m_1-1}^{(1, 2)}+e_{1, m_1-2}^{(1, 2)}\big)$.
Hence 
$\big[{q}_{\,m_1-1}^{(2,1)}, q_{\,m_1}^{(2,1)*}[2]\big]^\sharp\sim\big(2e_{2,1}^{(1, 1)}-e_{3,2}^{(1, 1)}+e_{m_1-1, m_1-2}^{(2, 2)}-2e_{m_1, m_1-1}^{(2, 2)}\big)^\sharp\sim(q_{\,2}^{(1, 1)}-q_{\,2}^{(2, 2)})$ and we obtain the element whose linear term is 
 \begin{align}\label{wt2 m1=m2}
 \omega\big(q_{\,2}^{(1,1)}\big)-\omega\big(q_{\,2}^{(2,2)}\big).
 \end{align}
\end{proof}

Consequently, Lemma \ref{lem:Thm-1} implies that the elements \eqref{weak generators} or \eqref{weak generators m_1=m_2} weakly generate the element whose linear term without the derivative $\partial$ is $c_1\omega\big(q_{\,2}^{(1, 1)}\big)+c_2\omega\big(q_{\,2}^{(2, 2)}\big)$ for some $c_1, c_2\in\CC$.
We use this element to prove (Step $2$) in the proof of Theorem \ref{big thm} and denote it by  $c_1\tilde{\omega}\big(q_{\,2}^{(1, 1)}\big)+c_2\tilde{\omega}\big(q_{\,2}^{(2, 2)}\big)$.

\begin{lem}[Step 2]\label{lem:Thm step2}
Let us consider the element  $c_1\tilde{\omega}\big(q_{\,2}^{(1, 1)}\big)+c_2\tilde{\omega}\big(q_{\,2}^{(2, 2)}\big)$.
\begin{enumerate}[(1)]
\item Suppose $m_1>m_2\geq2$. 
Then 
$\big(c_1\tilde{\omega}\big(q_{\,2}^{(1, 1)}\big)+c_2\tilde{\omega}\big(q_{\,2}^{(2, 2)}\big)\big)_{(2)}\omega\big(q_{\frac{m_1+m_2}{2}}^{(1, 2)}\big)\sim\omega\big(q_{\frac{m_1+m_2}{2}-1}^{(1, 2)}\big)$ 
and $\big(c_1\tilde{\omega}\big(q_{\,2}^{(1, 1)}\big)+c_2\tilde{\omega}\big(q_{\,2}^{(2, 2)}\big)\big)_{(2)}\omega\big(q_{\frac{m_1+m_2}{2}}^{(2, 1)}\big)\sim\omega\big(q_{\frac{m_1+m_2}{2}-1}^{(2, 1)}\big)$,
where $\sim$ is defined in \eqref{eq:sim}.
\item Suppose $m_1=m_2\geq2$.
Then $\big(c_1\tilde{\omega}\big(q_{\,2}^{(1, 1)}\big)+c_2\tilde{\omega}\big(q_{\,2}^{(2, 2)}\big)\big)_{(2)}\omega\big(q_{\,m_1}^{(1, 2)}\big)\sim \omega\big(q_{\,m_1-1}^{(1, 2)}\big)$. 
\end{enumerate}
\end{lem}
\begin{proof}
(1)
Recall from Lemma \ref{lem:Thm-1} that we have an element whose linear term is 
\begin{align}\label{wt2 diag12}
\omega\big(q_{\,2}^{(1, 1)}\big)+\frac{m_1(m_1+1)}{m_2(m_2+1)}\omega\big(q_{\,2}^{(2, 2)}\big)
\end{align} so that $c_1=1$ and $c_2=\frac{m_1(m_1+1)}{m_2(m_2+1)}$.
For every $b\in\g^f$, the coefficient of $\lambda^2$ in the $\lambda$-bracket $\{\tilde{\omega}\big(q_{\,2}^{(1, 1)}\big)\ { }_\lambda\ \omega(b)\}$ equals that in the following summation term of $\{\omega\big(q_{\,2}^{(1, 1)}\big)\ { }_\lambda\ \omega(b)\}$ using the formula in Theorem \ref{thm:main lemma},
\begin{equation}\label{last term of bracket formula}
        -\sum_{p= -1}^{N}\prod_{\substack{{t=0,\cdots, p+1},\\(j_{t-1}, n_{t-1})\prec(j_{t}, n_{t})}} \Big(\omega \big(\, \big[\, q_{j_{t-1}}[{n_{t-1}+1}]\, ,\,  q_{j_t}^*[n_t]\, \big]^\sharp\, \big)-( \, q_{j_{t-1}}[{n_{t-1}+1}]\, \big| \,  q_{j_t}^*[n_t]\, )\, (\lambda+\partial)\Big),
\end{equation}
for $N>\!>0$, $q_{\,2}^{(1, 1)}=q_{j_{p+1}}^*[n_{p+1}]$ and $b=q_{j_{-1}}[{n_{-1}+1}]$.
Since $(\ad e)^3q_{\,2}^{(1, 1)}=0$, the final factor of \eqref{last term of bracket formula} vanishes  
whenever  $n_p\geq2$ and hence corresponding terms of \eqref{last term of bracket formula} are zero in this case.  

Also, for $p\geq 1$ and $n_p=0$ and $n_{p-1}\geq 0$, the term with $t=p$ in the factor 
\begin{align*}
\omega \big(\, \big[\, q_{j_{p-1}}[{n_{p-1}+1}]\, ,\,  q_{j_p}^*[n_p]\, \big]^\sharp\, \big)-( \, q_{j_{p-1}}[{n_{p-1}+1}]\, \big| \,  q_{j_p}^*[n_p]\, )\, (\lambda+\partial)
\end{align*}
equals zero because $\big[\, q_{j_{p-1}}[{n_{p-1}+1}]\, ,\,  q_{j_p}^*[n_p]\, \big]^\sharp\sim[q_{j_{p-1}}, (\ad e)^{n_{p-1}+1} q_{j_p}^*]^\sharp=0$ and $( \, q_{j_{p-1}}[{n_{p-1}+1}]\, \big| \,  q_{j_p}^*[n_p]\, )\sim( \, q_{j_{p-1}}\, \big| \,  (\ad e)^{n_{p-1}+1}q_{j_p}^*\, )=0$. Thus all corresponding terms with $n_p=0$ and $n_{p-1}\geq0$ in  \eqref{last term of bracket formula} vanish. 
Therefore, the $\lambda^2$ term can appear only when $p=1, n_1=1$ and $n_0=0$ in \eqref{last term of bracket formula}.
Similarly, the coefficient of $\lambda^2$ in the $\lambda$-bracket $\{\tilde{\omega}\big(q_{\,2}^{(2, 2)}\big)\ { }_\lambda\ \omega(b)\}$ can be computed in the same way. 
Hence, if $b=q_{\frac{m_1+m_2}{2}}^{(1, 2)}$, then   
 \begin{align*}
\big(c_1\tilde{\omega}\big(q_{\,2}^{(1, 1)}\big)+c_2\tilde{\omega}\big(q_{\,2}^{(2, 2)}\big)\big)_{(2)}\omega\big(q_{\frac{m_1+m_2}{2}}^{(1, 2)}\big) \sim   \omega\Big(\,\big[q_{\frac{m_1+m_2}{2}}^{(1,2)},\, q_{\,2}^{(1,1)*}+\frac{m_1(m_1+1)}{m_2(m_2+1)}q_{\,2}^{(2,2)*}\big]^\sharp\,\Big).
 \end{align*}
Recall that $q_{\ 2}^{(1,1)*}\sim\sum\limits_{i=1}^{m_1-1}(im_1-i^2)e_{i, i+1}^{(1, 1)}$ and $q_{\ 2}^{(2,2)*}\sim\sum\limits_{j=1}^{m_2-1}(jm_2-j^2)e_{j, j+1}^{(2, 2)}$.
 As a result, $\big(c_1\tilde{\omega}\big(q_{\,2}^{(1, 1)}\big)+c_2\tilde{\omega}\big(q_{\,2}^{(2, 2)}\big)\big)_{(2)}\omega\big(q_{\frac{m_1+m_2}{2}}^{(1, 2)}\big)\sim \omega\big(q_{\frac{m_1+m_2}{2}-1}^{(1,2)}\big)$. 
 Similarly, we can obtain the lemma for $b=q_{\frac{m_1+m_2}{2}}^{(2, 1)}$.
 \\ 
 \nin(2)
 By analogous computations, we have $\big(c_1\tilde{\omega}\big(q_{\,2}^{(1, 1)}\big)+c_2\tilde{\omega}\big(q_{\,2}^{(2, 2)}\big)\big)_{(2)}\omega\big(q_{\,m_1}^{(1, 2)}\big)\sim\omega\big(\,\big[q_{\,m_1}^{(1,2)},\, q_{\,2}^{(1,1)*}-q_{\,2}^{(2,2)*}\big]^\sharp\,\big)$ and thus we get the lemma.
\end{proof}

By Lemma \ref{lem:Thm-1} and \ref{lem:Thm step2}, if $\omega\big(q_{\frac{m_1+m_2}{2}-1}^{(1, 2)}\big)$ and $\omega\big(q_{\frac{m_1+m_2}{2}-1}^{(2, 1)}\big)$ exist, then the elements \eqref{weak generators} or \eqref{weak generators m_1=m_2} weakly generate the elements $\omega\big(q_{\frac{m_1+m_2}{2}-1}^{(1, 2)}\big)$ and $\omega\big(q_{\frac{m_1+m_2}{2}-1}^{(2, 1)}\big)$.
We now prove that the elements \eqref{weak generators} or \eqref{weak generators m_1=m_2} weakly generate an element whose linear term is $\omega\big(q_{\,2}^{(1,1)}\big)$ up to a total derivative. 
In the case $m_1>m_2$ with $m_2=1$, we directly obtain the element having the linear term $\omega\big(q_{\,2}^{(1,1)}\big)$ from \eqref{wt2 diag12}.
For $m_1=m_2=2$, since $2m_1-5<0$, we use an alternative argument below. 
When $m_1=2$, $\omega\big(q_{\,m_1}^{(2,1)}\big)$ is obtained from the $0$th product $\eqref{wt2 m1=m2}_{(0)}\omega\big(q_{\,{m_1}-1}^{(2,1)}\big)$.
Consequently, the element with the linear term $\omega\big(q_{\,2}^{(1,1)}\big)$ can be expressed as a linear combination of \eqref{wt2 m1=m2} and $\omega\big(q_{\,m_1}^{(1,2)}\big) _{(2m_1-3)} \omega\big(q_{\,m_1}^{(2,1)}\big)$.
Hence, the lemma holds for all cases except these two.

\begin{lem}[Step 3]\label{lem:step3}
\begin{enumerate}[(1)]
\item Suppose $m_1>m_2\geq2$.
An element of the following form 
\begin{align}\label{wt2 diag}
\omega\big(q_{\,2}^{(1,1)}\big)+Q,
\end{align}
where $Q$ is a quadratic term in the finite set of variables $\{\,\omega(a_i)\,\}_{i=1}^n$ with $a_i\in\g^f_0$, can be obtained as a linear combination of $\omega\big(q_{\frac{m_1+m_2}{2}-1}^{(1, 2)}\big)_{(m_1+m_2-5)}\omega\big(q_{\frac{m_1+m_2}{2}-1}^{(2,1)}\big)$ and the element having \eqref{wt2 diag12}.
\item Suppose $m_1=m_2\geq3$. We obtain \eqref{wt2 diag} as a linear combination of $\omega\big(q_{\,{m_1}-1}^{(1,2)}\big) _{(2m_1-5)} \omega\big(q_{\,{m_1}-1}^{(2,1)}\big)$ and the element having \eqref{wt2 m1=m2}. 
\end{enumerate}
\end{lem}

\begin{proof}

(1)
As in the proof of Lemma \ref{lem:Thm-1}, the linear part without $\partial$ in $\omega\big(q_{\frac{m_1+m_2}{2}-1}^{(1, 2)}\big)_{(m_1+m_2-5)}\omega\big(q_{\frac{m_1+m_2}{2}-1}^{(2,1)}\big)$ is $
\omega\big(\big[{q}_{\frac{m_1+m_2}{2}-1}^{(2,1)}, q_{\frac{m_1+m_2}{2}-1}^{(2,1)*}[1]\big]^\sharp\big)$
up to a nontrivial constant multiple. 
Recall $q_{\frac{m_1+m_2}{2}-1}^{(2,1)*}[1]\sim\big(-(m_1-1)e_{1, m_2-2}^{(1, 2)}+(m_1-m_2)e_{2, m_2-1}^{(1, 2)}+(m_2-1)e_{3, m_2}^{(1, 2)}\ \big)$ and $q_{\frac{m_1+m_2}{2}-1}^{(2, 1)}=e_{m_2-1,1}^{(2, 1)}+e_{m_2, 2}^{(2, 1)}$.
Hence,
\begin{equation}\label{element for step3}
    \begin{aligned}
& \big[{q}_{\frac{m_1+m_2}{2}-1}^{(2,1)}, q_{\frac{m_1+m_2}{2}-1}^{(2,1)*}[1]\big]^\sharp \\
& \sim \quad \left(-(m_1-m_2)e_{2, 1}^{(1, 1)}-(m_2-1)e_{3, 2}^{(1, 1)}-(m_1-1)e_{m_2-1, m_2-2}^{(2, 2)}+(m_1-m_2)e_{m_2, m_2-1}^{(2,2)}\right)^\sharp.
    \end{aligned} 
\end{equation}
Thus, we obtain the element $\omega\big(q_{\,2}^{(1, 1)}\big)$ as a linear combination of $\omega\eqref{element for step3}$ and \eqref{wt2 diag12} and get the lemma. 
\\
\nin(2)
By using arguments similar to the proof of (1), the linear term of $\omega\big(q_{\,{m_1}-1}^{(1,2)}\big) _{(2m_1-5)} \omega\big(q_{\,{m_1}-1}^{(2,1)}\big)$ is 
 $\omega\big(q_{\,2}^{(1, 1)}\big)+ \omega\big(q_{\,2}^{(2, 2)}\big)$ up to scaling by a nonzero constant. 
 Thus, we have the element \eqref{wt2 diag} as a linear combination of \eqref{wt2 m1=m2} and $\omega\big(q_{\,{m_1}-1}^{(1,2)}\big) _{(2m_1-5)} \omega\big(q_{\,{m_1}-1}^{(2,1)}\big)$. 
\end{proof}

From the element \eqref{wt2 diag} obtained in the previous lemma, we generate homogeneous elements $\omega\big(q_{\,t}^{(1, 2)}\big)$ and $\omega\big(q_{\,t}^{(2, 1)}\big)$ for $1\leq t_{12}\leq m_2$ where $t_{jl}=t-\frac{|m_j-m_l|}{2}$, by repeated applications of the operator $\eqref{wt2 diag}_{(2)}$.

\begin{lem}[Step 4]\label{lem:Thm-2}
Let $1\leq t\leq m_2-2$ and denote the element \eqref{wt2 diag} by $\widetilde{\omega}\big( q_{\, 2}^{(1,1)}\big)$. Then the following statements hold.
\begin{enumerate}[(1)]
\item  $\widetilde{\omega}\big( q_{\, 2}^{(1,1)}\big)_{(2)}\omega\Big(q_{\frac{m_1+m_2}{2}-t}^{(1,2)}\Big)\sim  \omega\Big(q_{\frac{m_1+m_2}{2}-t-1}^{(1,2)}\Big)$.
\item $\widetilde{\omega}\big( q_{\, 2}^{(1,1)}\big)_{(2)}\omega\Big(q_{\frac{m_1+m_2}{2}-t}^{(2,1)}\Big)\sim \omega\Big(q_{\frac{m_1+m_2}{2}-t-1}^{(2,1)}\Big)$. 
\end{enumerate}
\end{lem}

\begin{proof}
We only prove the first statement since the second can be proved in the same way as the first. 
The coefficient of $\lambda^2$ in the $\lambda$-bracket $\big\{\widetilde{\omega}\big( q_{\, 2}^{(1,1)}\big)\ _\lambda\ \omega\big(q_{\frac{m_1+m_2}{2}-t}^{(1,2)}\big)\big\}$ equals that in the $\lambda$-bracket 
\begin{align}\label{bracket with q_2^11}
\Big\{\omega\big(q_{\,2}^{(1,1)}\big)\ _\lambda\ \omega\Big(q_{\frac{m_1+m_2}{2}-t}^{(1,2)}\Big)\Big\}.
\end{align}
By the proof of Lemma \ref{lem:Thm step2}, when applying the formula in Theorem \ref{thm:main lemma} to the bracket \eqref{bracket with q_2^11}, the $\lambda^2$ term occurs only in the summation term corresponding to $p=1, n_1=1$ and $n_0=0$. 
Hence, we have 
\begin{align*}
\widetilde{\omega}\big( q_{\, 2}^{(1,1)}\big)_{(2)}\omega\Big(q_{\frac{m_1+m_2}{2}-t}^{(1,2)}\Big)\sim\omega\Big(\,[q_{\frac{m_1+m_2}{2}-t}^{(1, 2)}, q_{\,2}^{(1, 1)*}]^\sharp\,\Big).
\end{align*}
Recall $q_{\frac{m_1+m_2}{2}-t}^{(1, 2)}=\sum\limits_{i=1}^{1+t}e^{(1, 2)}_{m_2-t+(i-1), i}$ and $q_{\,2}^{(1, 1)*}$ is in the proof of Lemma \ref{lem:Thm step2}.
Thus, $\big[q_{\frac{m_1+m_2}{2}-t}^{(1, 2)}\,, q_{\,2}^{(1, 1)*}\big]^\sharp\sim q_{\frac{m_1+m_2}{2}-t-1}^{(1, 2)}$ and then we get the lemma.
\end{proof}

In summary, for $m_1>m_2$, the elements $\omega\big(q_{\frac{m_1+m_2}{2}}^{(1,2)}\big)$ and $\omega\big(q_{\frac{m_1+m_2}{2}}^{(2,1)}\big)$ weakly generate, inductively, all elements $\omega\big(q_{\,t}^{(1, 2)}\big)$ and $\omega\big(q_{\,t}^{(2, 1)}\big)$ for $1\leq t_{12}\leq m_2$ where $t_{jl}=t-\frac{|m_j-m_l|}{2}$.
When $m_1=m_2$, the elements $\omega\big(q_{\,m_1}^{(1,2)}\big)$ and $\omega\big(q_{\,{m_1}-1}^{(2,1)}\big)$ weakly generate all such elements except for the element $\omega\big(q_{\,m_1}^{(2,1)}\big)$ analogously, and $\omega\big(q_{\,m_1}^{(2,1)}\big)$ is obtained from  $\widetilde{\omega}\big( q_{\, 2}^{(1,1)}\big)_{(0)}\omega\big(q_{\,{m_1}-1}^{(2,1)}\big)$.


\begin{lem}[Step 5]\label{lem:thm-3}
The elements \eqref{weak generators} or \eqref{weak generators m_1=m_2} weakly generate all elements whose linear term is of the form $\omega\big(q_{\,t}^{(j, l)}\big)$ with $j\neq l$ such that $(j,l)\neq (1, 2), (2, 1)$ and $1\leq t_{jl}\leq \text{min}\{m_j, m_l\}$.
\end{lem}
\begin{proof}
It is sufficient to prove that the elements \eqref{weak generators} or \eqref{weak generators m_1=m_2} weakly generate the elements whose linear term is of the form $\omega\big(q_{\,t}^{(3, 1)}\big)$ with $1\leq t_{13}\leq m_3$. 
By arguments similar to those in Lemma \ref{lem:Thm-1} and \ref{lem:step3}, it follows that there exists an element whose linear term is $\omega\big(q_{\,2}^{(j, j)}\big)$ for $j\geq2$, which we denote by $\widetilde{\omega}\big( q_{\, 2}^{(j,j)}\big)$ as in Lemma \ref{lem:Thm-2}.
Using Lemma \ref{lem:Thm-2} and a similar argument, the elements $\omega\big(q_{\frac{m_1-m_2}{2}+1}^{(2,1)}\big)$ and $\omega\big(q_{\frac{m_2-m_3}{2}+1}^{(3,2)}\big)$ are obtained from $\big(\widetilde{\omega}\big( q_{\, 2}^{(1,1)}\big)_{(2)}\big)^{m_2-2}\omega\Big(q_{\frac{m_1+m_2}{2}-1}^{(2,1)}\Big)$ and $\big(\widetilde{\omega}\big( q_{\, 2}^{(2,2)}\big)_{(2)}\big)^{m_3-1}\omega\Big(q_{\frac{m_2+m_3}{2}}^{(3,2)}\Big)$, respectively. 
If we denote by $A_{31}$ the linear term of 
\begin{align}\label{A31}
\omega\big(q_{\frac{m_1-m_2}{2}+1}^{(2,1)}\big)_{(0)}\omega\big(q_{\frac{m_2-m_3}{2}+1}^{(3,2)}\big),
\end{align}
then $A_{31}\sim\omega\big(q_{\frac{m_1-m_3}{2}+1}^{(3,1)}\big)$.
Hence, we can obtain an element whose linear term is $\omega\big(q_{\frac{m_1-m_3}{2}+t}^{(3,1)}\big)$ for $2\leq t\leq m_3$ from $\big(\widetilde{\omega}\big( q_{\, 2}^{(1,1)}\big)_{(0)}\big)^{t-1}\eqref{A31}$.
\end{proof}

The previous lemmas complete the proof for the non-diagonal blocks. 
It remains to deal with the diagonal blocks, which are stated in the following lemma. 

\begin{lem}[Step 6]\label{lem:Thm-4}
The elements \eqref{weak generators} or \eqref{weak generators m_1=m_2} weakly generate any element whose linear term is of the form $\omega
\big(q_{\,t}^{(j, j)}\big)$ for $1\leq j\leq d$ and $t=1$ or $3\leq t\leq m_j$.
\end{lem}

\begin{proof}
Suppose that $m_1>m_2$ and $t=1$ or $3\leq t\leq m_j$. We aim to show an element with the linear term $\omega\big(q_{\,t}^{(j, j)}\big)$ can be obtained as a linear combination of $\omega\big(q_{\frac{m_j+m_{j+1}}{2}}^{(j, j+1)}\big)_{(m_j+m_{j+1}-t-1)}\omega\big(q_{\frac{m_j+m_{j+1}}{2}}^{(j+1, j)}\big)$ and $\omega\big(q_{\frac{m_j+m_{j+1}}{2}-1}^{(j, j+1)}\big)_{(m_j+m_{j+1}-t-3)}\omega\big(q_{\frac{m_j+m_{j+1}}{2}-1}^{(j+1, j)}\big)$.

Let us show the claim for $j=1$. For $j>1,$ we omit the proof since it can be proved by the same argument.
Following the proof of Lemma \ref{lem:Thm-1}, from $\omega\big(q_{\frac{m_1+m_2}{2}}^{(1,2)}\big)_{(m_1+m_2-t-1)} \omega\big(q_{\frac{m_1+m_2}{2}}^{(2,1)}\big)$, we get the elements whose linear term is
$\omega\big(\,[q_{\frac{m_1+m_2}{2}}^{(2,1)}, q_{\frac{m_1+m_2}{2}}^{(2,1)*}[t-1]\,]^\sharp\big)$ for $1\leq t\leq m_2$.
When $t=1$, we have $\big[q_{\frac{m_1+m_2}{2}}^{(2,1)}, q_{\frac{m_1+m_2}{2}}^{(2,1)*}\big]^\sharp\sim\big[e_{m_2, 1}^{(2, 1)}, e_{1, m_2}^{(1, 2)}\big]^\sharp\sim q_{\,1}^{(2,2)}$. 
Next, assume $3\leq t\leq m_2$ and consider  
\begin{align}\label{large wt get}
\big[q_{\frac{m_1+m_2}{2}}^{(2,1)}, q_{\frac{m_1+m_2}{2}}^{(2,1)*}[t-1]\big]^\sharp\sim\big(-e_{t, 1}^{(1, 1)}+(-1)^{t-1}e_{m_2, m_2-t+1}^{(2, 2)}\, \big)^\sharp.
\end{align} 
Consequently, we have an element whose linear term is $\omega\eqref{large wt get}$ from $\omega\big(q_{\frac{m_1+m_2}{2}}^{(1,2)}\big)_{(m_1+m_2-t-1)} \omega\big(q_{\frac{m_1+m_2}{2}}^{(2,1)}\big)$.
Likewise, from $\omega\big(q_{\frac{m_1+m_2}{2}-1}^{(1, 2)}\big)_{(m_1+m_2-t-3)}\omega\big(q_{\frac{m_1+m_2}{2}-1}^{(2, 1)}\big)$, we get the elements whose linear term is 
\begin{align*}
\omega\Big(\,\big[q_{\frac{m_1+m_2}{2}-1}^{(2,1)}, q_{\frac{m_1+m_2}{2}-1}^{(2,1)*}[t-1]\big]^\sharp\,\Big).
\end{align*}
Since we know $\big[q_{\frac{m_1+m_2}{2}-1}^{(2,1)}, q_{\frac{m_1+m_2}{2}-1}^{(2,1)*}[t-1]\big]$ is constant multiple of 
\begin{align}\label{second large wt get}
 \begin{split}
&-\Big(\, \big((m_1-1)-(m_2-1)(t-1)\big)e_{t, 1}^{(1, 1)}+(m_2-1)e_{t+1, 2}^{(1, 1)}\Big)
\\&+(-1)^{t-1}\Big((m_1-1)e_{m_2-1, m_2-t}^{(2, 2)}+(-(t-1)(m_1-1)+m_2-1)e_{m_2, m_2-t+1}^{(2, 2)}\Big),
 \end{split}
\end{align}
we get an element whose linear term is $\omega\big(\eqref{second large wt get}^\sharp\big)$.
By direct computations, we obtain the elements whose linear term is $\omega\big(q_{\, t}^{(2,2)}\big)$ as a linear combination of $\omega\eqref{large wt get}$ and $\omega\big(\eqref{second large wt get}^\sharp\big)$.

Suppose that $m_1=m_2$. 
Unlike the case $m_1>m_2$, for $t=1$ or $3\leq t\leq m_2$,
the element with the linear term $\omega\big(q_{\,t}^{(2, 2)}\big)$ is obtained from the linear combination of
\begin{align*}
\omega\big(q_{\,m_1}^{(1,2)}\big)_{(2m_1-t-2)}\omega\big(q_{\,{m_1}-1}^{(2,1)}\big)\ \mathrm{and}\ \omega\big(q_{\,m_1-1}^{(1,2)}\big)_{(2m_1-t-3)}\omega\big(q_{\,{m_1}-1}^{(2,1)}\big).
\end{align*} 
In particular, when $m_1=m_2=2$, the element with the linear term $\omega\big(q_{\,t}^{(2, 2)}\big)$ is obtained from the linear combination of $\omega\big(q_{\,m_1}^{(1,2)}\big)_{(2m_1-t-2)}\omega\big(q_{\,{m_1}-1}^{(2,1)}\big)$ and $\omega\big(q_{\,m_1}^{(1,2)}\big)_{(2m_1-t-1)}\omega\big(q_{\,{m_1}}^{(2,1)}\big)$ for $t=1$.
\end{proof}

Combining Lemma \ref{lem:Thm-1}--\ref{lem:Thm-4}, we complete the proof of Theorem \ref{big thm}. 
We now turn to general case where $\g=\mathfrak{sl}_{N_1|N_2}$ with $N_1\neq N_2$ and let $f$ be an even nilpotent element whose Jordan block is of type $(m_1, m_2, \cdots, m_d)$. 
Here, we assume that 
(i)\,$N_1=m_1+\cdots+m_{d_1}\,, \, m_1\geq\cdots\geq m_{d_1}$(ii)\,$N_2=m_{d_1+1}+\cdots+m_{d_1+d_2}\,,\,m_{d_1+1}\geq\cdots\geq m_{d_1+d_2}\,$(iii)\,$d=d_1+d_2\geq2$.
In this setting, we obtain the super analogue of Theorem \ref{big thm}. 
Since its proof is essentially the same as that of Theorem \ref{big thm}, we focus only on the differences below.

\begin{thm}\label{big thm super}
Let $\g=\mathfrak{sl}_{N_1|N_2}$ and $f$ be its nilpotent element described as above.
The $2d-2$ elements of the same form as those in Theorem \ref{big thm} weakly generates $\mathcal{W}^k(\g, f)$.
\end{thm} 
\begin{proof}
When $d_1\geq 2$, the proof is identical to that for the Lie algebra case. 
In contrast, if $d_1=1$ and $m_1\neq m_2$, the linear part \eqref{wt2 diag12} is replaced by 
\begin{align}\label{wt2 diag12 parity} 
\omega\big(q_{\,2}^{(1,1)}\big)-\frac{m_1(m_1+1)}{m_2(m_2+1)}\omega\big(q_{\,2}^{(2,2)}\big),
\end{align}
and this substitution leaves the rest of the proof unchanged.
In the remaining case where $d_1=1$ and $m_1=m_2$, 
elements $\omega\big(q_{\,m_1}^{(1,2)}\big) _{(2m_1-4)} \omega\big(q_{\,m_1-1}^{(2,1)}\big)$ and $\omega\big(q_{\frac{m_2+m_3}{2}}^{(2,3)}\big) _{(m_2+m_3-3)} \omega\big(q_{\frac{m_2+m_3}{2}}^{(3,2)}\big)$ produce elements whose linear terms are $\omega\big(q_{\,2}^{(1,1)}\big)+\omega\big(q_{\,2}^{(2,2)}\big)$ and $\omega\big(q_{\,2}^{(2,2)}\big)-\frac{m_1(m_1+1)}{m_3(m_3+1)}\omega\big(q_{\,2}^{(3,3)}\big)$, respectively.
Consequently, the proof remains unchanged with the linear term \eqref{wt2 m1=m2} replaced by the element $\omega\big(q_{\,2}^{(1,1)}\big)+\frac{m_1(m_1+1)}{m_3(m_3+1)}\omega\big(q_{\,2}^{(3,3)}\big)$.
 \end{proof}

\subsection{Weak generating set of $\mathcal{W}^k(\g,f)$ with small conformal Weights}\label{Chap:small}
In Section \ref{Subsec:large}, we found weak generating sets with large conformal weights and, to get other strong generators, higher order poles of weak generators are needed as in (Step 6). 
In the section, we construct a weak generating set consisting of elements with small conformal weights and use lower order poles to get other strong generators.
The following theorem is the main result of this section.

\begin{thm}\label{big thm small}
Let $\g$ and $f$ be given as Theorem \ref{big thm} and recall the Notation \eqref{notation}. 
\begin{enumerate}[(1)]
\item If $m_1>m_2$, then the following $2d-1$ elements
\begin{align}\label{weak generators small}
\omega\big(q_{\,3}^{(1,1)}\big)\,,\, \omega\Big(q_{\frac{m_i-m_{i+1}}{2}+1}^{(i,i+1)}\Big)\,,\, \omega\Big(q_{\frac{m_i-m_{i+1}}{2}+1}^{(i+1,i)}\Big)\, (1\leq i\leq d-1),
\end{align}
weakly generates $\mathcal{W}^k(\g,f)$ for $k\neq 0$, 
 except in the case of $m_1=2$. When $m_1=2$, we include  $\omega\big(q_{\,2}^{(1,1)}\big)$ instead of $\omega\big(q_{\,3}^{(1,1)}\big)$.
\item If $m_1=m_2$, then the following $2d-1$ elements 
\begin{align}\label{weak generators m_1=m_2 small} 
\omega\big(q_{\,3}^{(1,1)}\big)\,,\, \omega\big(q_{\,2}^{(2,1)}\big)\,,\, \omega\Big(q_{\frac{m_i-m_{i+1}}{2}+1}^{(i,i+1)}\Big)\,,\, \omega\Big(q_{\frac{m_j-m_{j+1}}{2}+1}^{(j+1,j)}\Big)\, (1\leq i\leq d-1, \ 2\leq j\leq d-1),
\end{align}
weakly generates $\mathcal{W}^k(\g,f)$ for $k\neq 0$, except in the case of $m_1=2$. When $m_1=2$, we include  $\omega\big(q_{\,2}^{(1,1)}\big)$ instead of $\omega\big(q_{\,3}^{(1,1)}\big)$.
\end{enumerate}
\end{thm}

\vskip 3mm 

\noindent \textbf{Outline of the Proof}   

\begin{enumerate}[\ (\text{Step}1)]
\item Get an element having $\omega\big(q_{\,2}^{(1, 1)}\big)$ :
\begin{itemize}
\item Get this element from $\omega\big(q_{\,3}^{(1,1)}\big)_{(3)}\omega\big(q_{\,3}^{(1,1)}\big)$. 
\end{itemize}
\item Get an element having $\omega\big(q_{\frac{\lvert m_j-m_l\rvert}{2}+1}^{(j, l)}\big)$  for $j\neq l$ :
\begin{itemize}
\item If $m_1=m_2$, get an element with $\omega\big(q_{\,1}^{(2, 1)}\big)$ from $\omega\big(q_{\,2}^{(1, 1)}\big)_{(2)}\omega\big(q_{\,2}^{(2, 1)}\big)$.
\item  If $m_1>m_2$, we already have $\omega\big(q_{\frac{m_1-m_2}{2}+1}^{(2, 1)}\big)$.
\item Consider $\omega\big(q_{\frac{m_1-m_2}{2}+1}^{(2, 1)}\big)_{(0)}\omega\big(q_{\frac{m_2-m_3}{2}+1}^{(3, 2)}\big)$ to get an element with $\omega\big(q_{\frac{m_1-m_3}{2}+1}^{(3, 1)}\big)$.
\item All other generators can be obtained in a similar way. 
\end{itemize}

\item Get an element having $\omega\big(q_{\,1}^{(j,j)}\big)$ or $\omega\big(q_{2}^{(j, j)}\big)$ for $2\leq j\leq d$ :
\begin{itemize}
\item Get $\omega\big(q_{\,1}^{(2,2)}\big)$ from $\omega\big(q_{\frac{m_1-m_2}{2}+1}^{(1,2)}\big)_{(m_1-m_2)}\omega\big(q_{\frac{m_1-m_2}{2}+1}^{(2,1)}\big)$. 
\item If $m_1>m_2$, get $d_1\omega\big(q_{\,2}^{(1, 1)}\big)+d_2\omega\big(q_{\,2}^{(2, 2)}\big)$ for $d_1, d_2\in\mathbb{C}$
 from $\omega\big(q_{\frac{m_1-m_2}{2}+1}^{(1, 2)}\big)_{(m_1-m_2-1)}\omega\big(q_{\frac{m_1-m_2}{2}+1}^{(2, 1)}\big)$.
 \item If $m_1=m_2$, get $d_1\omega\big(q_{\,2}^{(1, 1)}\big)+d_2\omega\big(q_{\,2}^{(2, 2)}\big)$ for $d_1, d_2\in\mathbb{C}$ from $\omega\big(q_{\frac{m_1-m_2}{2}+1}^{(1, 2)}\big)_{(m_1-m_2)}\omega\big(q_{\frac{m_1-m_2}{2}+2}^{(2, 1)}\big)$.
 \item Combine with the element in (Step1).
\item All other generators can be obtained in a similar way. 
\end{itemize}
\item Get an element having $\omega\big(q_{\,3}^{(j,j)}\big)$ for $2\leq j\leq d$ :
\begin{itemize}
 \item If $m_1\geq m_2+2$, get $c_1\omega\big(q_{\,3}^{(1, 1)}\big)+c_2\omega\big(q_{\,3}^{(2, 2)}\big)$ for $c_1, c_2\in\mathbb{C}$ from  $\omega\big(q_{\frac{m_1-m_2}{2}+1}^{(1,2)}\big)_{(m_1-m_2-2)}\omega\big(q_{\frac{m_1-m_2}{2}+1}^{(2,1)}\big)$. 
 \item If $m_1=m_2+1$, get $c_1\omega\big(q_{\,3}^{(1, 1)}\big)+c_2\omega\big(q_{\,3}^{(2, 2)}\big)$ for $c_1, c_2\in \mathbb{C}$ from  
$\big(\omega\big(q_{\,2}^{(1,1)}\big)_{(0)}\omega\big(q_{\frac{m_1-m_2}{2}+1}^{(1,2)}\big)\big)_{(0)}\omega\big(q_{\frac{m_1-m_2}{2}+1}^{(2,1)}\big)$.  
\item If $m_1=m_2$, get $c_1\omega\big(q_{\,3}^{(1, 1)}\big)+c_2\omega\big(q_{\,3}^{(2, 2)}\big)$ for $c_1, c_2\in\mathbb{C}$ from $\big(\omega\big(q_{\,2}^{(1,1)}\big)_{(0)}\omega\big(q_{\,1}^{(1,2)}\big)\big)_{(0)}\omega\big(q_{\,2}^{(2,1)}\big)$.
\item All other generators can be obtained in a similar way. 
\end{itemize}
\item Get an element having $\omega\big(q_{\,t}^{(j, l)}\big)$ for $2\leq t_{jl}\leq \text{min}\{m_j, m_l\}$ where $t_{jl}=t-\frac{\lvert m_j-m_l\rvert}{2}$ :
\begin{itemize}
\item If $2\leq t_{jl}\leq\text{min}\{m_j, m_l\}$, consider $\omega\big(q_{\,3}^{(j,j)}\big)_{(1)}\omega\big(q_{\,t-1}^{(j, l)}\big)$ or $\omega\big(q_{\,2}^{(j, j)}\big)_{(0)}\omega\big(q_{\,t-1}^{(j,l)}\big)$ to get an element with $\omega\big(q_{\, t}^{(j, l)}\big)$.
\end{itemize} 
\end{enumerate}
Once the initial conditions have been constructed through (Step $1$--$4$) in the proof, all the strong generators can be obtained inductively by using only the $0$-th and first products. 
The argument through (Step $1$--$5$) leads to a strong generating set of $\mathcal{W}^k(\g, f)$, which proves the theorem. \qed

\vskip 3mm 

Now, we provide the detailed proof of each step in Lemma \ref{lem:step1 small}--\ref{lem:step5 small}, based on Theorem \ref{thm:main lemma}.

\begin{lem}[Step 1]\label{lem:step1 small}
Suppose $m_1\geq 3$.
Let $A$ be the linear term of $\omega\big(q_{\,3}^{(1,1)}\big)_{(3)}\omega\big(q_{\,3}^{(1,1)}\big)$ without the total derivative part. 
Then $A\sim\omega\big(q_{\,2}^{(1, 1)}\big)$.
\end{lem}
\begin{proof}
Notice that $q_{\,3}^{(1,1)}\sim q_{\,3}^{(1,1)*}[4]$.
Then as in the proof of Lemma \ref{lem:Thm-1} , to obtain the linear term of $\omega\big(q_{\,3}^{(1,1)}\big)_{(3)}\omega\big(q_{\,3}^{(1,1)}\big)$, it suffices to consider the term
\begin{align*}
\big[\, q_{\,3}^{(1, 1)}\, ,\,  q_{\,3}^{(1, 1)*}[1]\, \big]^\sharp.
\end{align*}
Recall that $q_{\,3}^{(1, 1)}=\sum\limits_{j=1}^{m_1-2}e_{j+2, j}^{(1, 1)}$ and $q_{\,3}^{(1, 1)*}[1]\sim\sum\limits_{j=1}^{m_1-1}j(m_1-j)(2j-m_1)e_{j, j+1}^{(1, 1)}$.
Therefore, the desired linear term is obtained, and the lemma is proved. 
\end{proof}

To show that the elements \eqref{weak generators small} or \eqref{weak generators m_1=m_2 small} weakly generate all elements with  $\omega\big(\,q_{\frac{\lvert m_j-m_l\rvert}{2}+1}^{(j, l)}\big)$ where $j\neq l$, it suffices to prove the following lemma.

\begin{lem}[Step 2]\label{lem:step2 small}
Let $1\leq l<n<j\leq d$.
\begin{enumerate}[(1)]
\item Suppose $m_1=m_2$.
Then $\omega\big(q_{\,2}^{(1, 1)}\big)_{(2)}\omega\big(q_{\,2}^{(2, 1)}\big)\sim \omega\big(q_{\,1}^{(2, 1)}\big)$ .
\item
Let $B$ denote the linear term of $\omega\big(q_{\frac{m_l-m_n}{2}+1}^{(n,l)}\big)_{(0)}\omega\big(q_{\frac{m_n-m_j}{2}+1}^{(j,n)}\big)$ without the total derivative part.
Then we obtain $B\sim\omega\big(q_{\frac{m_l-m_j}{2}+1}^{(j,l)}\big)$. 
\end{enumerate}
\end{lem}
\begin{proof}
(1)
By a similar argument to the proof of Lemma \ref{lem:step3}, we obtain $\omega\big(q_{\,2}^{(1, 1)}\big)_{(2)}\omega\big(q_{\,2}^{(2, 1)}\big)\sim\omega\big(\big[q_{\,2}^{(2, 1)}, q_{\,2}^{(1, 1)*}\big]^\sharp\big)$.
Recall that $q_{\,2}^{(2,1)}=\sum\limits_{j=1}^{m_1-1}e_{j+1, j}^{(2, 1)}$ and $q_{\,2}^{(1,1)*}\sim\sum\limits_{i=1}^{m_1-1}(im_1-i^2)e_{i, i+1}^{(1, 1)}$.
Thus we have $\big[q_{\,2}^{(2, 1)}, q_{\,2}^{(1, 1)*}\big]^\sharp\sim q_{\,1}^{(2, 1)}$ and hence we get $\omega\big(q_{\,2}^{(1, 1)}\big)_{(2)}\omega\big(q_{\,2}^{(2, 1)}\big)\sim \omega\big(q_{\,1}^{(2, 1)}\big)$. 
\\
\nin(2)
It follows from the formula in Theorem \ref{thm:main lemma} that the linear term of $\omega(a)_{(0)}\omega(b)$ is $\omega\big([a, b]\big)$ up to a total derivative part if $a, b\in\g^f$.
Without loss of generality, for the case $l=1, n=2$ and $j=3$, it is enough to show that $\big[q_{\frac{m_l-m_n}{2}+1}^{(n,l)}\, , \, q_{\frac{m_n-m_j}{2}+1}^{(j,n)}\big]\sim q_{\frac{m_l-m_j}{2}+1}^{(j,l)}\,$.
Recall that $q_{\frac{m_1-m_2}{2}+1}^{(2,1)}=\displaystyle\sum\limits_{j=1}^{m_2}e_{j, j}^{(2, 1)}$ and $q_{\frac{m_2-m_3}{2}+1}^{(3,2)}=\displaystyle\sum\limits_{j=1}^{m_3}e_{j, j}^{(3, 2)}$.
Consequently, $\big[q_{\frac{m_1-m_2}{2}+1}^{(2,1)}\,,\, q_{\frac{m_2-m_3}{2}+1}^{(3,2)}\big]=-q_{\frac{m_1-m_3}{2}+1}^{(3,1)}$ and so we get the lemma.
\end{proof}

Using the elements obtained in the above lemma whose linear term is of the form $\omega\big(\,q_{\frac{\lvert m_j-m_l\rvert}{2}+1}^{(j, l)}\big)$ where $j\neq l$, we show how to obtain the elements in the diagonal block with conformal weights $1$ and $2$.

\begin{lem}[Step 3]\label{small wt get}
Let $1\leq j<d$.
\begin{enumerate}[(1)]
\item $\omega\big(q_{\frac{m_j-m_{j+1}}{2}+1}^{(j,j+1)}\big)_{(m_j-m_{j+1})}\omega\big(q_{\frac{m_j-m_{j+1}}{2}+1}^{(j+1,j)}\big)\sim \big(b_1\omega\big(q_{\,1}^{(j,j)}\big)+b_2\omega\big(q_{\,1}^{(j+1,j+1)}\big)\big)$ for some $b_1, b_2\in\CC^\times$.
\item Suppose $m_j>m_{j+1}$.
The linear term of $\omega\big(q_{\frac{m_j-m_{j+1}}{2}+1}^{(j,j+1)}\big)_{(m_j-m_{j+1}-1)}\omega\big(q_{\frac{m_j-m_{j+1}}{2}+1}^{(j+1,j)}\big)$ is $d_1\omega\big(q_{\,2}^{(j, j)}\big)+d_2\omega\big(q_{\,2}^{(j+1, j+1)}\big)$ for some $d_1, d_2\in\mathbb{C}^\times$, up to a total derivative part.
\item Suppose $m_j=m_{j+1}$. 
The linear term of $\omega\big(q_{\frac{m_j-m_{j+1}}{2}+1}^{(j,j+1)}\big)_{(0)}\omega\big(q_{\frac{m_j-m_{j+1}}{2}+2}^{(j+1,j)}\big)$ is $d_1\omega\big(q_{\,2}^{(j, j)}\big)+d_2\omega\big(q_{\,2}^{(j+1, j+1)}\big)$ for some $d_1, d_2\in\mathbb{C}^\times,$ up to a total derivative part.
\end{enumerate}
\end{lem}

\begin{proof}
(1)
Thanks to the skew-symmetry of the $\lambda$-bracket, we have $\omega\big(q_{\frac{m_1-m_2}{2}+1}^{(1,2)}\big)_{(m_1-m_2)}\omega\big(q_{\frac{m_1-m_2}{2}+1}^{(2,1)}\big)\sim \omega\big(q_{\frac{m_1-m_2}{2}+1}^{(2,1)}\big)_{(m_1-m_2)}\omega\big(q_{\frac{m_1-m_2}{2}+1}^{(1,2)}\big)$ and  it suffices to prove that $\big(q_{\frac{m_1-m_2}{2}+1}^{(2,1)}\big)_{(m_1-m_2)}\omega\big(q_{\frac{m_1-m_2}{2}+1}^{(1,2)}\big)\sim\omega\big(q_{\,1}^{(2,2)}\big)$.
Recall that $q_{\frac{m_1-m_2}{2}+1}^{(2,1)}\sim q_{\frac{m_1-m_2}{2}+1}^{(1,2)*}[m_1-m_2]$.
Then, following the argument used in the proof of Lemma \ref{lem:Thm-1}, to obtain the linear term of $\omega\big(q_{\frac{m_1-m_2}{2}+1}^{(2,1)}\big)_{(m_1-m_2)}\omega\big(q_{\frac{m_1-m_2}{2}+1}^{(1,2)}\big)$, it is enough to compute $
\omega\big(\big[\, q_{\frac{m_1-m_2}{2}+1}^{(1, 2)}\, ,\,  q_{\frac{m_1-m_2}{2}+1}^{(1, 2)*}\big]^\sharp\big)$.
Recall that $q_{\frac{m_1-m_2}{2}+1}^{(1,2)}=\sum\limits_{j=1}^{m_2}e_{m_1-m_2+j, j}^{(1, 2)}$ and $q_{\frac{m_1-m_2}{2}+1}^{(1, 2)*}\sim\sum\limits_{j=1}^{m_2}{{m_1-m_2+j-1 \choose j-1 }}e_{j, m_1-m_2+j}^{(2, 1)}$.
Hence, $\big[q_{\frac{m_1-m_2}{2}+1}^{(1, 2)}\, ,\,  q_{\frac{m_1-m_2}{2}+1}^{(1, 2)*} \big]^\sharp\sim q_{\,1}^{(2,2)}$ and the claim follows. 
\\ \nin(2)
As in the proof of (1), it suffices to compute the term $
\big[\, q_{\frac{m_1-m_2}{2}+1}^{(1, 2)}\, ,\,  q_{\frac{m_1-m_2}{2}+1}^{(1, 2)*}[1]\, \big]^\sharp$ for $m_1>m_2$ case.
Recall that $q_{\frac{m_1-m_2}{2}+1}^{(1,2)}=\sum\limits_{j=1}^{m_2}e_{m_1-m_2+j, j}^{(1, 2)}$ and $q_{\frac{m_1-m_2}{2}+1}^{(1, 2)*}[1]\sim\sum\limits_{j=1}^{m_2}{{m_1-m_2+j-2 \choose j-1 }}e_{j, m_1-m_2+j-1}^{(2, 1)}$.
Thus 
\begin{equation*}
    \begin{aligned}
& \big[\, q_{\frac{m_1-m_2}{2}+1}^{(1, 2)}\, ,\,  q_{\frac{m_1-m_2}{2}+1}^{(1, 2)*}[1]\, \big]^\sharp\\
& \sim\Big(\sum\limits_{j=1}^{m_2}{{m_1-m_2+j-2 \choose j-1 }}e_{m_1-m_2+j, m_1-m_2+j-1}^{(1, 1)}-\sum\limits_{j=2}^{m_2}{{m_1-m_2+j-2 \choose j-1 }}e_{j, j-1}^{(2, 2)}\Big)^\sharp.
\end{aligned}
\end{equation*}
Hence, the linear term of $\omega\big(q_{\frac{m_1-m_2}{2}+1}^{(1,2)}\big)_{(m_1-m_2-1)}\omega\big(q_{\frac{m_1-m_2}{2}+1}^{(2,1)}\big)$ produces $\omega\big(q_{\,2}^{(1, 1)}\big)-\frac{m_1(m_1-1)}{m_2(m_2+1)}\omega\big(q_{\,2}^{(2, 2)}\big)$ up to a total derivative part. 
When $m_1=2$, we have $m_2=1$ and the linear term of $\omega\big(q_{\frac{m_1-m_2}{2}+1}^{(1,2)}\big)_{(m_1-m_2-1)}\omega\big(q_{\frac{m_1-m_2}{2}+1}^{(2,1)}\big)$ induces $\omega\big(q_{\,2}^{(1, 1)}\big)$ up to a total derivative part. 
\\
\nin(3)
Suppose $m_1=m_2$. 
Recall that if $a\in\g^f_0$, then $\omega(a)_{(0)}\omega(b)=\omega\big([a, b]\big)$ for $b\in\g^f$.
Since $q_{\,1}^{(1,2)}\in\g^f_0$, we have
\begin{align*}
\omega\big(q_{\,1}^{(1,2)}\big)_{(0)}\omega\big(q_{\,2}^{(2,1)}\big)=\omega\Big(\big[q_{\,1}^{(1,2)}\,,\,q_{\,2}^{(2,1)}\big]\Big)=\omega\big(q_{\,2}^{(1,1)}\big)-\omega\big(q_{\,2}^{(2,2)}\big)
\end{align*}
which proves the lemma.
\end{proof}

 
Elements of Higher conformal weights are obtained recursively via the first product with an element of the conformal weight $3$. For this purpose, we need an element with the linear term $\omega\big(q_{\,3}^{(j, j)}\big)$. In the following lemma, we show how to get this element.

\begin{lem}[Step 4]\label{lem:step4 small}
Let $j\geq2$ be an index such that $m_j\geq 3$.
\begin{enumerate}[(1)]
\item Suppose $m_{j-1}\geq m_j+2$. 
The linear term of $\omega\big(q_{\frac{m_{j-1}-m_j}{2}+1}^{(j-1,j)}\big)_{(m_{j-1}-m_j-2)}\omega\big(q_{\frac{m_{j-1}-m_j}{2}+1}^{(j,j-1)}\big)$ is $c_1\omega\big(q_{\,3}^{(j-1, j-1)}\big)+c_2\omega\big(q_{\,3}^{(j, j)}\big)$ for some $c_1, c_2\in \mathbb{C}^\times$ up to a total derivative part.
\item Suppose $m_{j-1}=m_j+1$. 
The linear term of $\big(\omega\big(q_{\,2}^{(j-1,j-1)}\big)_{(0)}\omega\big(q_{\frac{m_{j-1}-m_j}{2}+1}^{(j-1,j)}\big)\big)_{(0)}\omega\big(q_{\frac{m_{j-1}-m_j}{2}+1}^{(j,j-1)}\big)$ is $c_1\omega\big(q_{\,3}^{(j-1, j-1)}\big)+c_2\omega\big(q_{\,3}^{(j, j)}\big)$ for some $c_1\in\mathbb{C}$ and $c_2\in \mathbb{C}^\times$ up to a total derivative part.
\item Suppose $m_{j-1}=m_j$. 
The linear term of $\big(\omega\big(q_{\,2}^{(j-1,j-1)}\big)_{(0)}\omega\big(q_{\frac{m_{j-1}-m_j}{2}+1}^{(j-1,j)}\big)\big)_{(0)}\omega\big(q_{\frac{m_{j-1}-m_j}{2}+2}^{(j,j-1)}\big)$ is $c_1\omega\big(q_{\,3}^{(j-1, j-1)}\big)+c_2\omega\big(q_{\,3}^{(j, j)}\big)$ for some $c_1, c_2\in\mathbb{C}^\times,$ up to a total derivative part.
\end{enumerate} 
\end{lem} 

\begin{proof}
If $m_2<3$, then there is nothing to prove. 
Thus, we assume $m_2\geq 3$.
\\
\nin(1)
Without loss of generality, it suffices to prove that the linear term of $\omega\big(q_{\frac{m_1-m_2}{2}+1}^{(1,2)}\big)_{(m_1-m_2-2)}\omega\big(q_{\frac{m_1-m_2}{2}+1}^{(2,1)}\big)$ is of the form $c_1\omega\big(q_{\,3}^{(1, 1)}\big)+c_2\omega\big(q_{\,3}^{(2, 2)}\big)$ for some $c_1, c_2\in\mathbb{C}^\times$ and $m_1\geq m_2+2$, up to a total derivative part.
Following the argument in the proof of Lemma \ref{small wt get}, the linear term of the above expression is
\begin{align*}
\omega\Big(\big[\, q_{\frac{m_1-m_2}{2}+1}^{(1, 2)}\, ,\,  q_{\frac{m_1-m_2}{2}+1}^{(1, 2)*}[2]\, \big]^\sharp\Big),
\end{align*}
up to a nontrivial constant multiple. 
Recall that $q_{\frac{m_1-m_2}{2}+1}^{(1, 2)*}[2]\sim\sum\limits_{j=1}^{m_2}{{m_1-m_2+j-3 \choose j-1 }}e_{j, m_1-m_2+j-2}^{(2, 1)}$ and $q_{\frac{m_1-m_2}{2}+1}^{(1,2)}=\sum\limits_{j=1}^{m_2}e_{m_1-m_2+j, j}^{(1, 2)}$.
Therefore,$\big[\, q_{\frac{m_1-m_2}{2}+1}^{(1, 2)}\, ,\,  q_{\frac{m_1-m_2}{2}+1}^{(1, 2)*}[2]\, \big]^\sharp\sim\big( q_{\,3}^{(1, 1)}-\frac{m_1(m_1-1)(m_1-2)}{m_2(m_2+1)(m_2+2)}q_{\,3}^{(2, 2)}\big)$ and the claim follows.
\\
\nin(2)
Next, it suffices to show that if $m_1=m_2+1$, then the linear term of $\big(\omega\big(q_{\,2}^{(1,1)}\big)_{(0)}\omega\big(q_{\frac{3}{2}}^{(1,2)}\big)\big)_{(0)}\omega\big(q_{\frac{3}{2}}^{(2,1)}\big)$ is of the form $c_1\omega\big(q_{\,3}^{(1, 1)}\big)+c_2\omega\big(q_{\,3}^{(2, 2)}\big)$ for some $c_1\in\CC$ and $ c_2\in \mathbb{C}^\times,$ up to a total derivative part.
Let $A_{12}$ denote the linear term of $\omega\big(q_{\,2}^{(1,1)}\big)_{(0)}\omega\big(q_{\frac{3}{2}}^{(1,2)}\big)$  without total derivatives.
As in the proof of Lemma \ref{lem:thm-3}, $A_{12}\sim\omega\big(\big[\,q_{\frac{3}{2}}^{(1,2)}, q_{\,2}^{(1,1)*}[2]\big]^\sharp\big)$, where $q_{\,2}^{(1, 1)*}[2]\sim\sum\limits_{j=1}^{m_1-1}e_{j+1, j}^{(1, 1)}$ and thus we have $A_{12}\sim\omega\big(q_{\frac{5}{2}}^{(1,2)}\big)$. 
Analogously to the proof of Lemma \ref{small wt get}, the linear term of $\big(\omega\big(q_{\,2}^{(1,1)}\big)_{(0)}\omega\big(q_{\frac{3}{2}}^{(1,2)}\big)\big)_{(0)}\omega\big(q_{\frac{3}{2}}^{(2,1)}\big)$ equals 
\begin{align*}
\omega\Big(\big[q_{\frac{5}{2}}^{(1, 2)}\, ,\,  q_{\frac{5}{2}}^{(1, 2)*}[2]\big]^\sharp\Big),
\end{align*}
up to a nontrivial constant multiple. 
Recall that $q_{\frac{5}{2}}^{(1,2)}=\sum\limits_{j=1}^{m_2-1}e_{j+2, j}^{(1, 2)}$ and $q_{\frac{5}{2}}^{(1, 2)*}[2]\sim\sum\limits_{j=1}^{m_2}(2m_2-6j+4)e_{j, j}^{(2, 1)}$.
Since $\big(\sum\limits_{j=1}^{m_2-2}(2m_2-6j-8)e_{j+2, j}^{(2, 2)}\big)^\sharp\sim q_{\,3}^{(2, 2)}$, this completes the second part of the proof. 
\\
\nin(3)
Finally, suppose $m_1=m_2$.
It remains to show that the linear term of $\big(\omega\big(q_{\,2}^{(1,1)}\big)_{(0)}\omega\big(q_{\,1}^{(1,2)}\big)\big)_{(0)}\omega\big(q_{\,2}^{(2,1)}\big)$ is of the form $c_1\omega\big(q_{\,3}^{(1, 1)}\big)+c_2\omega\big(q_{\,3}^{(2, 2)}\big)$ for some $c_1, c_2\in \mathbb{C}^\times,$ up to a total derivative part.
As in Lemma \ref{lem:step2 small}, $\omega\big(q_{\,2}^{(1,1)}\big)_{(0)}\omega\big(q_{\,1}^{(1,2)}\big)\sim\omega\big(q_{\,2}^{(1,2)}\big)$.
Since the linear term of $\omega(a)_{(0)}\omega(b)$ is $\omega\big([a, b]\big)$ up to a total derivative part, the linear term of $\omega\big(q_{\,2}^{(1,2)}\big)_{(0)}\omega\big(q_{\,2}^{(2,1)}\big)$ is 
\begin{align*}
\omega\Big(\big[q_{\,2}^{(1,2)}, q_{\,2}^{(2,1)}\big]\Big)
\end{align*}
up to a total derivative part.
Recall that $q_{\,2}^{(1,2)}=\sum\limits_{j=1}^{m_1-1}e_{j+1, j}^{(1, 2)}$ and $q_{\,2}^{(2,1)}=\sum\limits_{j=1}^{m_1-1}e_{j+1, j}^{(2, 1)}$.
Hence $\big[q_{\,2}^{(1,2)}, q_{\,2}^{(2,1)}\big]=q_{\,3}^{(1,1)}-q_{\,3}^{(2,2)}$ and therefore the linear term of $\big(\omega\big(q_{\,2}^{(1,1)}\big)_{(0)}\omega\big(q_{\,1}^{(1,2)}\big)\big)_{(0)}\omega\big(q_{\,2}^{(2,1)}\big)$ induces an element with the linear term $\omega\big(q_{\,3}^{(1,1)}\big)-\omega\big(q_{\,3}^{(2,2)}\big)$.
\end{proof}

Thus, Lemma \ref{lem:step4 small} shows that whenever the element $q_{\,3}^{(j, j)}$ exists, the element whose linear term is $\omega\big(q_{\,3}^{(j, j)}\big)$, up to a total derivative part, can be weakly generated by the elements \eqref{weak generators small} or \eqref{weak generators m_1=m_2 small}.
Finally, we state the lemma describing how to weakly generate elements whose linear term is of the form $\omega\big(q_{\,t}^{(j, l)}\big)$ with $t\geq\frac{\lvert m_j-m_l\rvert}{2}+2$.

\begin{lem}[Step 5]\label{lem:step5 small}
Let $j$ be the index such that $m_j\geq 3$ and let $j_1$ denote the index satisfying $1\leq j_1\leq d$.
\begin{enumerate}[(1)]
\item For $1+\delta_{m_j, m_l}\leq t\leq\mathrm{min}\{m_j, m_l\}-1$,
the linear term of $\omega\big(q_{\,3}^{(j,j)}\big)_{(1)}\omega\big(q_{\frac{\lvert m_j-m_l\rvert}{2}+t}^{(j, l)}\big)$ is $\omega\big(\,q_{\frac{\lvert m_j-m_l\rvert}{2}+t+1}^{(j, l)}\big)$ and that of  $\omega\big(q_{\,3}^{(j,j)}\big)_{(1)}\omega\big(\,q_{\frac{\lvert m_j-m_l\rvert}{2}+t}^{(l, j)}\big)$ is $\omega\big(q_{\frac{\lvert m_j-m_l\rvert}{2}+t+1}^{(l, j)}\big)$, up to a nontrivial constant multiple and up to a total derivative part. 
\item For $1\leq t\leq\mathrm{min}\{m_{j_1}, m_l\}-1$, the linear term of $\omega\big(q_{\,2}^{(j_1,j_1)}\big)_{(0)}\omega\big(q_{\frac{\lvert m_{j_1}-m_l\rvert}{2}+t}^{(j_1, l)}\big)$ is $\omega\big(q_{\frac{\lvert m_{j_1}-m_l\rvert}{2}+t+1}^{(j_1, l)}\big)$ and that of $\omega\big(q_{\,2}^{(j_1,j_1)}\big)_{(0)}\omega\big(q_{\frac{\lvert m_{j_1}-m_l\rvert}{2}+t}^{(l, j_1)}\big)$ is $\omega\big(q_{\frac{\lvert m_{j_1}-m_l\rvert}{2}+t+1}^{(l, j_1)}\big)$, up to a nontrivial constant multiple and up to a total derivative part. 
\end{enumerate}
\end{lem}

\begin{proof}
(1) From the proof of Lemma \ref{lem:step1 small}, one can see that $q_{\ 3}^{(j,j)}\sim q_{\ 3}^{(j,j)*}[4]$ and $q_{\ 3}^{(j,j)*}[1] \sim \sum_{i=1}^{m_j-1}i(m_j-1)(2i-m_j) e_{i\, i+1}^{(j,j)}.$ Hence we get $[f,[f, q_{\ 3}^{(j,j)*}[1]]= q_{\ 3}^{(j,j)*}[3]\sim \sum_{i=1}^{m_j-1} i \,  e_{i+1,i}^{(j,j)}.$ Finally, since the linear term of $\omega\big(q_{\,3}^{(j,j)}\big)_{(1)}\omega\big(q_{\frac{\lvert m_j-m_l\rvert}{2}+t}^{(j, l)}\big)$ is the constant multiple of 
\[ \omega\Big(\Big[ \, q_{\frac{\lvert m_j-m_l\rvert}{2}+t}^{(j, l)}\, ,\, q^{(j,j)*}_{\, 3}[3] \, \Big]^\sharp \Big)\sim \omega \Big( q_{\frac{\lvert m_j-m_l\rvert}{2}+t+1}^{(j, l)} \Big),\]
we get the first assertion of the lemma.
Also, the same proof works for $\omega\big(q_{\,3}^{(j,j)}\big)_{(1)}\omega\big(\,q_{\frac{\lvert m_j-m_l\rvert}{2}+t}^{(l, j)}\big)$. \\
(2) directly follows from the fact that $\omega(a)_{(0)}\omega(b)= \omega{[a,b]}$ for any $a,b\in \g^f.$
\end{proof}

Lemma \ref{lem:step1 small}--\ref{lem:step5 small} complete the proof of Theorem \ref{big thm small}, as strong generators are constructed from the elements listed in the theorem.
We now turn to the general case where $\g=\mathfrak{sl}_{N_1|N_2}$ with $N_1\neq N_2$ and $f$ is an even nilpotent element whose Jordan block is of type $(m_1, m_2, \cdots, m_d)$, satisfying
(i)\,$N_1=m_1+\cdots+m_{d_1}\,, \, m_1\geq\cdots\geq m_{d_1}$(ii)\,$N_2=m_{d_1+1}+\cdots+m_{d_1+d_2}\,,\,m_{d_1+1}\geq\cdots\geq m_{d_1+d_2}$(iii)\,$d=d_1+d_2\geq2$.
The generalization of Theorem \ref{big thm small} is stated below. 
Its proof is almost the same as that of Theorem \ref{big thm small}, with just minor modifications analogous to those appearing in the proof of Theorem \ref{big thm super}. 
Hence, we omit the details.  

\begin{thm}\label{big thm small super}
Let $\g= \mathfrak{sl}_{N_1|N_2}$ and $f$ be its nilpotent element  described as above.
\begin{enumerate}[(1)]
\item If max$\{m_1, m_{d_1+1}\}=2$, then the $2d-1$
elements of the same form as those in Theorem \ref{big thm small} weakly generate $\mathcal{W}^k(\g, f)$, except that in each element, the term $m_i-m_{i+1}$ is replaced by $\lvert m_i-m_{i+1}\rvert$. Also, $\omega\big(q_{\,2}^{(1, 1)}\big)$ is replaced by $\omega\big(q_{\,2}^{(l, l)}\big)$
\item 
Suppose max$\{m_1, m_{d_1+1}\}\geq3$ and choose $l$ such that max$\{m_1, m_{d_1+1}\}=m_l$. 
Then the $2d-1$ elements of the same form as those in Theorem \ref{big thm small} weakly generate $\mathcal{W}^k(\g, f)$, except that $\omega\big(q_{\,3}^{(1, 1)}\big)$ is replaced by $\omega\big(q_{\,3}^{(l, l)}\big)$ and in the elements containing the term $m_i-m_{i+1}$, this term is replaced by its absolute value $\lvert m_i-m_{i+1}\rvert$.
\end{enumerate}
\end{thm} 

\subsection{Weak generators of quantum W-algebras}

In this section, we deduce properties of weak generating sets of quantum W-algebras by using those of classical W-algebras, together with the relationship between quantum and classical W-algebras described in Section \ref{subsec:Q-C W-algebra}. We investigate relations between quantum and classical W-algebras to derive properties of weak generating sets of quantum W-algebras.

Recall the complex $C(\g,f,k)_{\epsilon}$ in \eqref{eq:complex-with parameter} with the differential $Q^\epsilon:= \frac{1}{\epsilon}d^\epsilon_{k\, (0)}$ for the elements $d^\epsilon_k$ defined in \eqref{element d-epsilon}. Let us consider $k$ and $\epsilon$ as formal variables and define a grading $\text{gr}_{\epsilon}$ on $\mathbb{C}[\lambda]\otimes C(\g,f,k)_{\epsilon}$ as follows:
\begin{equation} \label{eq:grading-epsilon}
    \text{gr}_{\epsilon}(\partial^n A)=\text{gr}_\epsilon(\lambda^n)=n, \quad \text{gr}_{\epsilon}(k)=\text{gr}_\epsilon(\epsilon)=-1, \quad \text{gr}_{\epsilon}(:AB:)=\text{gr}_\epsilon(A)+ \text{gr}_\epsilon(B),
\end{equation}
for $A,B\in C(\g,f,k)_{\epsilon}.$ Then the following lemma holds:

\begin{lem}
The grading $\text{gr}_{\epsilon}$ is well-defined. More precisely, we have 
\begin{enumerate}[(1)]
    \item $\text{gr}_\epsilon(:AB:)=\text{gr}_{\epsilon}(:BA:),$
    \item $\text{gr}_\epsilon(:A:BC::)=\text{gr}_{\epsilon}(::AB:C:),$
    \item $\text{gr}_{\epsilon}([A{}_\lambda B])=\text{gr}_\epsilon(A)+\text{gr}_\epsilon(B)-1,$
\end{enumerate}
for homogeneous elements $A,B,C\in C(\g,f,k)_{\epsilon}.$
\end{lem}
\begin{proof}
    For a given ordered basis $\mathcal{B}$ of $r:=\g\oplus \phi_\n \oplus \phi^{\n^*}\oplus \Phi_{\g_{1/2}},$ one can take a PBW basis with respect to the ordered basis $\bigsqcup_{i \in \mathbb{Z}_+} \partial^i \mathcal{B}$ of $\mathbb{C}[\partial]\otimes r$. Consider the monomial grading $\text{gr}_M$ defined by $\text{gr}_M(:\partial^{n_1} u_{i_1} \cdots \partial^{n_t} u_{i_t}:)=t$ for each basis element in the PBW basis. Now we prove the lemma using the induction on the  grading $\text{gr}_M.$ To check the initial condition, let $a, b\in r.$ One can show (3) for $A=\partial^t a$ and $B=\partial^s b$ directly from the $\lambda$-bracket \eqref{eq:bracket_epsilon_second} and the sesqui-linearity. Also, (1) is true from the quasi-commutativity and the fact that 
    \[ \text{gr}_\epsilon\Big(\int^0_{-\partial} [A{}_\lambda B] d\lambda \Big)= 1+  \text{gr}_\epsilon( [A{}_\lambda B])=\text{gr}_\epsilon(A)+\text{gr}_{\epsilon}(B). \]
    Hence the lemma is true when $\text{gr}_M(A)+\text{gr}_M(B)+\text{gr}_M(C)=2.$ Suppose the lemma is true when $\text{gr}_M(A)+\text{gr}_M(B)+\text{gr}_M(C)=n.$ In the case when $\text{gr}_M(A)+\text{gr}_M(B)+\text{gr}_M(C)=n+1,$ the property (3) follows from the induction hypothesis and the Wick formula, that is 
    \[ \text{gr}_{\epsilon}\Big(\int^\lambda_0 [[A{}_\lambda B]{}_\mu C] d\mu\Big)= \text{gr}_\epsilon(A)+\text{gr}_{\epsilon}(B)+\text{gr}_{\epsilon}(C)-1.\]
    Similarly, (1) and (2) hold by the induction hypothesis and the quasi-commutativity and quasi-associativity, respectively.
\end{proof}

\begin{cor}
    The complex $C(\g,f,k)_{\epsilon}$ can be decomposed as 
    \[ C(\g,f,k)_{\epsilon}=\bigoplus_{i\in \mathbb{Z}}C_{\epsilon}^{[i]},\]
    where $C_{\epsilon}^{[i]}$ is the degree $i$ subspace of $C(\g,f,k)_{\epsilon}$ with respect to $\text{gr}
    _\epsilon$ and the differential $Q^\epsilon$ is the degree preserving differential, that is 
    \[ Q^\epsilon(C_{\epsilon}^{[i]})\subset C_{\epsilon}^{[i]}, \text{ for } i\in \mathbb{Z}.\]
\end{cor}

Now let us observe the grading on classical W-algebra which is induced from $\text{gr}_\epsilon$. Recall the Poisson bracket \eqref{eq:bracket_classical} as a quasi-classical limit of a vertex algebra over $\mathbb{C}[\epsilon].$ Consequently, the complex $\mathcal{C}(\g,f,k)$ in \eqref{eq:classical_complex} has the induced grading from $\text{gr}_\epsilon$ defined by 
\begin{equation} \label{eq:grading_classical}
    \text{gr}_{\epsilon}(\partial^n A)=n,\quad \text{gr}_\epsilon(k)=1, \quad \text{gr}_\epsilon(AB)= \text{gr}_\epsilon(A) + \text{gr}_\epsilon(B),
\end{equation}
for $A,B\in \mathcal{C}(\g,f,k)$. In addition, when we consider the Poisson $\lambda$-bracket, we extend the grading by $\text{gr}_\epsilon(\lambda)=1.$ One can check that $\text{gr}_\epsilon(\{A{}_\lambda B\})=\text{gr}_\epsilon(A)+\text{gr}_\epsilon(B).$
It is well-known that the isomorphism between the classical W-algebra in Definition \ref{def:classical W} and $W^k(\g,f)_{\epsilon}|_{\epsilon=0}$ is induced from the differential algebra homomorphism 
\begin{equation} \label{eq:corr, two classical}
    \mathcal{V}(\mathfrak{p}) \to \mathcal{C}(\g,f,k), \quad a\mapsto J_a,
\end{equation}
where $J_a= a+\sum_{\alpha\in S_{>0}} :\phi^\alpha \phi_{[u_\alpha, a]}$ for $a\in \bigoplus_{i\leq 0}\g_i$ and $J_a=-\Phi_a$ for $a\in \g_{1/2}.$ Since $\text{gr}_\epsilon(J_a)=0,$ the grading \eqref{eq:grading_classical} can be understood as a grading on $\mathcal{V}(\mathfrak{p})$ when $A,B\in \mathcal{V}(\mathfrak{p}).$

\begin{lem} \label{eq:level K W-alg}
    Recall $\mathcal{Q}=d^{cl}_{k\, (0)}$ is the differential in the classical BRST $\mathcal{C}(\g,f,k)$ 
    and let us denote by $\omega_k$ the map \eqref{eq:omega} to clarify it is the map to the level $k$ classical W-algebra. 
    \begin{enumerate}[(1)]
        \item $\mathcal{Q}$ preserved the grading $\text{gr}_\epsilon.$
        \item For any $a\in \g^f,$ the element $\text{gr}_\epsilon(\omega_k(\g,f))=0.$
    \end{enumerate}
\end{lem}
\begin{proof}
    (1) follows from the fact that $\text{gr}_\epsilon(d^{cl}_{k\, (0)})=0$ and $\text{gr}_{\epsilon}(\{A{}_\lambda B\})= \text{gr}_\epsilon(A)+\text{gr}_{\epsilon}(B).$ To see (2), notice that for any element in $\mathcal{W}^k(\g,f),$ each homogeneous component is again an element in $\mathcal{W}^k(\g,f).$ Now since $\omega_k(a)$ has the leading term $a$ which has degree $0$ part, with respect to $\text{gr}_\epsilon,$ the degree $0$-part of $\omega^0_k(a)$ is also a nontrivial element in  $\mathcal{W}(\g,f).$ However, $\omega^0_k(a)$ has also the properties (i) and (ii) below \eqref{eq:omega} and hence the uniqueness of such element implies $\omega^0_k(a)=\omega_k(a).$
\end{proof}

As in Section \ref{subsec:Q-C W-algebra}, we can view $\mathcal{C}(\g,f,k)$ as the $\epsilon^0$-part of $C(\g,f,k)$ and $\mathcal{W}^k(\g,f)=W^k(\g,f)_\epsilon|_{\epsilon=0}.$ Hence there is an element 
\begin{equation}
    J_{\{a\}}^\epsilon= J_{\{a\}}^{0} + \sum_{i\geq 1}\epsilon^i J_{\{a\}}^i   \in C(\g,f,k)_{\epsilon}, 
\end{equation}
satisfying the following properties:
\begin{enumerate}[(i)]
    \item $J_{\{a\}}^i\in C(\g,f,k)_{\epsilon}|_{\epsilon=0}$ and $J_{\{a\}}^{0}$ is the image of $\omega_k(a)$ by the map \eqref{eq:corr, two classical}, 
    \item $J_{\{a\}}^{\epsilon=1}=\sum_{i\in \mathbb{Z}_+}J_{\{a\}}^i$ is in the quantum W-algebra $W^k(\g,f),$
    \item $\text{gr}_\epsilon(J_{\{a\}}^\epsilon)=0$, that is $\text{gr}_\epsilon(J_{\{a\}}^i)=i$ for $i\in \mathbb{Z}_+.$ 
\end{enumerate}

\vskip 2mm 

In the following lemma, we observe $n$-th products between two elements in $W^k(\g,f).$ It is well known that $W^k(\g,f)$ is freely generated by $\{J^{\epsilon=1}_{\{a_i\}}| i\in S^f\}$ where $\{a_i|i\in S^f\}$ is a basis of $\g^f$. Hence we call $\partial ^n J^{\epsilon=1}_{\{a\}}$, for $n\in \mathbb{Z}_+$ and $a\in \g^f$, a linear element in $W^k(\g,f)$.

\begin{lem}\label{lem:weak generator same number}
    Let $a,b\in \g^f$ and $k\in \mathbb{C}$ be a nonzero element.
    Suppose $\omega_k(a)_{(n)}\omega_k(b)\neq 0$ and the linear term of $\omega_k(a)_{(n)}\omega_k(b)$ has a nontrivial constant multiple of $\omega(a_i)$ for $i\in S^f.$ Then, for generic values of $k\in \mathbb{C},$ the linear term of $J_{\{a\}}^{\epsilon=1}{}_{(n)}J_{\{b\}}^{\epsilon=1}$ has the nontrivial constant multiple of $ J^{\epsilon=1}_{\{a_i\}}$.
\end{lem}
\begin{proof}
    For the moment, let us consider $k$ as a formal variable to use the grading $\text{gr}_\epsilon$.
    As mentioned above, we have $\text{gr}_\epsilon(J^\epsilon_{\{a\}})=\text{gr}_\epsilon(J^\epsilon_{\{b\}})=0$ and hence $\text{gr}_\epsilon(J^\epsilon_{\{a\}}{}_{(n)}J^\epsilon_{\{b\}})= -n-1.$ This implies 
    \[ \text{gr}_{\epsilon}(\epsilon^i J^i_{\{a\}}{}_{(n)}\epsilon^j J^j_{\{b\}})=-n-1\]
    for any $i,j\in \mathbb{Z}_+.$ By the assumption and the grading $\text{gr}_\epsilon$ consideration, $J^0_{\{a\}}{}_{(n)}J^0_{\{b\}}$ has the nontrivial constant multiple of $\epsilon k^{n} J^0_{\{a_i\}}$. If $i$ or $j$ is strictly bigger than $0$, $\epsilon k^{n} J^0_{\{a_i\}}$ cannot arise from $\epsilon^i J^i_{\{a\}}{}_{(n)}\epsilon^j J^j_{\{b\}}$, since it should be in $\epsilon^{i+j+1}C(\g,f,k)_{\epsilon}.$ Hence in $W^k(\g,f)$, the $n$-th product $J_{\{a\}}^{\epsilon=1}{}_{(n)}J_{\{b\}}^{\epsilon=1}$ has the nontrivial constant multiple of $k^n J^{\epsilon=1}_{\{a_i\}}$ if and only if $\frac{1}{\epsilon}(J^0_{\{a\}}{}_{(n)} J^0_{\{b\}})|_{\epsilon=0}$ has. It is also equivalent to say that the linear term of $\omega_k(a)_{(n)}\omega_k(b)$ has the nontrivial constant multiple of $k^n\omega(a_i)$ for $i\in S^f.$ 
    
    However, $J_{\{a\}}^{\epsilon=1}{}_{(n)}J_{\{b\}}^{\epsilon=1}$ can involve constant multiples of $k^m J^{\epsilon=1}_{\{a_i\}}$ for $m<n.$ Consider the polynomial $f(k)$ such that    $J_{\{a\}}^{\epsilon=1}{}_{(n)}J_{\{b\}}^{\epsilon=1}-f(k)J^{\epsilon=1}_{\{a_i\}}$ has no term with $k^m J^{\epsilon=1}_{\{a_i\}}$ for any $m\in \mathbb{Z}$. Then $J_{\{a\}}^{\epsilon=1}{}_{(n)}J_{\{b\}}^{\epsilon=1}$ has no term with $J^{\epsilon=1}_{\{a_i\}}$ only when $k$ satisfies $f(k)=0.$ Since $f(k)$ has only finitely many zeros, one can conclude that the assertion holds for generic $k.$ 
\end{proof}

Recall in Section \ref{Chap:large}, we tracked only linear part in $n$-th product between generators $\omega(a)$ and $\omega(b)$ for $a,b\in \g^f.$ Hence we get the following theorem as a direct consequence of Lemma \ref{lem:weak generator same number}.

\begin{thm} \label{thm:weak generator-quantum case}
Let $\g=\mathfrak{sl}_N$ or $\mathfrak{sl}_{N_1|N_2}$ and let $f$ be an even nilpotent element of $\g$ in Theorem $\ref{big thm}$ or $\ref{big thm super}$.
Let $m_l=\text{max}\{m_1, m_{d_1+1}\}$.
\begin{enumerate}[(1)]
\item If $m_1\neq m_2$, then the following $2d-2$ elements 
\begin{align*}
J_{\{q_{\frac{m_{i}+m_{i+1}}{2}}^{(i,i+1)}\}}^{\epsilon=1}\,,\,  J_{\{q_{\frac{m_{i}+m_{i+1}}{2}}^{(i+1,i)}\}}^{\epsilon=1} \ \text{ for } \   1\leq i\leq d-1
\end{align*}
weakly generate $W^{k}(\g, f)$ for generic values of $k\in\CC$. Also, the following $2d-1$ elements
\begin{align*}
J_{\{q_{\,3}^{(l,l)}\}}^{\epsilon=1}\,,\,J_{\{q_{\frac{\lvert m_{i}-m_{i+1}\rvert}{2}+1}^{(i,i+1)}\}}^{\epsilon=1}\,,\,  J_{\{q_{\frac{\lvert m_{i}-m_{i+1}\rvert}{2}+1}^{(i+1,i)}\}}^{\epsilon=1} \ \text{ for } \  1\leq i\leq d-1
\end{align*}
weakly generate $W^{k}(\g, f)$ for generic values of $k\in\CC$, except in the case of $m_l=2$. When $m_l=2$, we include  $J_{\{q_{\,2}^{(l,l)}\}}^{\epsilon=1}$ instead of $J_{\{q_{\,3}^{(l,l)}\}}^{\epsilon=1}$.

\item If $m_1=m_2$, then the following $2d-2$ elements 
\begin{align*}
J_{\{q_{\, m_1-1}^{(2,1)}\}}^{\epsilon=1}\,,\, J_{\{q_{\frac{m_{i}+m_{i+1}}{2}}^{(i,i+1)}\}}^{\epsilon=1}\,,\,  J_{\{q_{\frac{m_{j}+m_{j+1}}{2}}^{(j+1,j)}\}}^\epsilon\,  (1\leq i\leq d-1,\ 2\leq j\leq d-1)
\end{align*}
weakly generate $W^{k}(\g, f)$ for generic values of $k\in\CC$. Also, the following $2d-1$ elements
\begin{align*}
J_{\{q_{\,3}^{(l,l)}\}}^{\epsilon=1}\,,\,J_{\{q_{\,2}^{(2,1)}\}}^{\epsilon=1}\,,\, J_{\{q_{\frac{\lvert m_{i}-m_{i+1}\rvert}{2}+1}^{(i,i+1)}\}}^{\epsilon=1}\,,\,  J_{\{q_{\frac{\lvert m_{j}-m_{j+1}\rvert}{2}+1}^{(j+1,j)}\}}^{\epsilon=1}\,(1\leq i\leq d-1,\ 2\leq j\leq d-1)
\end{align*}
weakly generate $W^{k}(\g, f)$ for generic values of $k\in\CC$, except in the case of $m_l=2$.  When $m_l=2$, we include  $J_{\{q_{\,2}^{(l,l)}\}}^{\epsilon=1}$ instead of $J_{\{q_{\,3}^{(l,l)}\}}^{\epsilon=1}$. 
\end{enumerate}
\end{thm}

\begin{proof}
By Theorem \ref{big thm}, \ref{big thm super}, \ref{big thm small} and \ref{big thm small super}, the classical W-algebra $\mathcal{W}^k(\g, f)$ is weakly generated by $M=2d-1$ elements $\{\omega(a_{i_j})|j=1,\cdots, M\}$ for some $i_j\in S^f$. 
It follows that any element $\omega(a_i)$ for $i\in S^f$ can be obtained as the linear term of a linear combination of elements of the form $\omega(a_{i_j})_{(n)}\omega(a_{i_l})$ with $n\geq0$.
By Lemma \ref{lem:weak generator same number}, for generic values of $k\in\CC$, the corresponding element $J_{\{a_i\}}^{\epsilon=1}$ appears as a linear term of $J_{\{a_{i_j}\}}^{\epsilon=1}{}_{(n)}J_{\{a_{i_l}\}}^{\epsilon=1}$.
Therefore, $W^k(\g,f)$ is weakly generated by $\{J_{\{a_{i_j}\}}^{\epsilon=1}|\,{i_j}\in S^f\,,\,j=1, \cdots, M\}$ for generic values of $k\in\CC$.
\end{proof}

\begin{rem}
Theorem \ref{thm:weak generator-quantum case} is weak generation analogue of Theorem \ref{explicit form of generator of affine}. However, it remains an open question how to characterize the levels for which the theorem holds. 
\end{rem}

\section{Examples}

When $\g= \mathfrak{sl}_N$ and $f$ is its principal nilpotent, we described weak generators of $\mathcal{W}^k(\mathfrak{sl}_N,f)$ in 
Example \ref{ex:principal-1}. 
In this section, we extend our investigation to weak generators of several other classical $W$-algebras, based on the results in Section \ref{Chap:large}. Additionally, we briefly outline the procedure for obtaining strong generators from these weak generators
in each case.

\subsection{$\mathfrak{sl}_{n+1|n}$}\label{Subsec:n+1}
Let $\g=\mathfrak{sl}_{n+1|n}$ and $f$ be even nilpotent element whose Jordan block is of type $(n+1, n)$.
Then the nilpotent element $f$ of type $(n+1, n)$ is $f=\sum\limits_{i=1}^{n}e_{i+1, i}^{(1, 1)}+\sum\limits_{j=1}^{n-1}e_{j+1, j}^{(2, 2)}.$
The basis of $\g^f$ in Lemma \ref{lem:basis g^f} is  \[\mathcal{B}^f=\{q_{\, t_1}^{(1,1)}\,,\, q_{\,l+\frac{1}{2}}^{(1, 2)}\,,\, q_{\,l+\frac{1}{2}}^{(2, 1)}\,,\,q^{(2, 2)}_{\, t_2}\,|\,2\leq t_1\leq n+1\,,\, 1\leq l\leq n\,,\, 1\leq t_2\leq n  \,\}.\] 
By Theorem \ref{big thm super}, the set \[
\mathcal{C}^f_{big}=\Big\{\omega\big(q_{\,n+\frac{1}{2}}^{(1,2)}\big)\,,\, \omega\big(q_{\,n+\frac{1}{2}}^{(2,1)}\big)\Big\}\] weakly generates $\mathcal{W}^k(\g,f).$  A weight two generator $\ell_1:=\omega\big(q_{\,n+\frac{1}{2}}^{(1,2)}\big)_{(2n-2)}\omega\big(q_{\,n+\frac{1}{2}}^{(1,2)}\big)$ can be used to get the lower weight elements $\omega\big(q_{\,n-\frac{1}{2}}^{(1,2)}\big)$ and $\omega\big(q_{\,n-\frac{1}{2}}^{(2,1)}\big)$. In addition, we get another weight 2 strong generator $\ell_2:=\omega\big(q_{\,n-\frac{1}{2}}^{(1,2)}\big)_{(2n-4)}\omega\big(q_{\,n-\frac{1}{2}}^{(1,2)}\big)$. Inductively, other strong generators of the form $\omega(q^{(1,2)}_{l-1/2})$ or $\omega(q^{(1,2)}_{l-1/2})$ can be obtained by $\ell_{(2)}\omega(q^{(1,2)}_{l+1/2})$ or $\ell_{(2)}\omega(q^{(1,2)}_{l+1/2})$, where $\ell$ is a linear combination of $\ell_1$ and $\ell_2$. 
Finally, we can get the strong generators $\omega(q_{t_i}^{(i,i)})$ for $i=1,2$ by computing the two OPEs $\{\omega\big(q_{\,n+\frac{1}{2}}^{(1,2)}\big){}_\lambda \omega(q_{\,n+\frac{1}{2}}^{(1,2)})\}$ and $\{\omega\big(q_{\,n-\frac{1}{2}}^{(1,2)}\big){}_\lambda \omega(q_{\,n-\frac{1}{2}}^{(1,2)})\}$. 
Suppose $n\geq 2$. Another weak generating set introduced in Theorem \ref{big thm small super} is 
\[\mathcal{C}^f_{sm}=\Big\{\omega\big(q_{\ \frac{3}{2}}^{(1,2)}\big)\,,\, \omega\big(q_{\ \frac{3}{2}}^{(2,1)}\big)\,,\, \omega\big(q_{\ 3}^{(1,1)}\big)\Big\}.\]
From $\mathcal{C}^f_{sm}$, we get strong generators $\omega\big(q_{\ t}^{(i,i)}\big)$ for $t\leq 3$ and $i=1,2$
by computing the OPEs between the generators in $\mathcal{C}^f_{sm}$.  After that, we inductively get higher weight strong generators $\omega(q_t^{(i,j)})$ by considering $\omega(q_3^{(i,i)})_{(1)}\omega(q_{t-1}^{(i,j)})$ for $i,j=1,2.$

\subsection{$\mathfrak{sl}_{m|n}$ with $m\geq n+2$}\label{Subsec:n+2}
Let $\g=\mathfrak{sl}_{m|n}$ with $m\geq n+2$ and $f$ be even nilpotent element whose Jordan block is of type $(m, n)$.
Then the element $f$ is identified with $f=\sum\limits_{i=1}^{m-1}e_{i+1, i}^{(1, 1)}+\sum\limits_{j=1}^{n-1}e_{j+1, j}^{(2, 2)}$ and 
\begin{align*}
\mathcal{B}^f=\{q_{\, t_1}^{(1,1)}\,,\, q_{\,l+\frac{m-n}{2}}^{(1, 2)}\,,\, q_{\,l+\frac{m-n}{2}}^{(2, 1)}\,,\,q^{(2, 2)}_{\, t_2}\,|\,2\leq t_1\leq m\,,\, 1\leq l\leq n\,,\, 1\leq t_2\leq n  \,\}.
\end{align*}
By Theorem \ref{big thm super}, $\mathcal{C}^f_{big}=\big\{\omega\big(q_{\frac{m+n}{2}}^{(1,2)}\big)\ ,\ \omega\big(q_{\frac{m+n}{2}}^{(2,1)}\big)\big\}$ weakly generates $\mathcal{W}^k(\g, f)$.
On the other hand, by Theorem \ref{big thm small super}, the set $
\mathcal{C}^f_{sm}=\big\{\,\omega\big(q_{\frac{m-n}{2}+1}^{(1,2)}\big)\,,\,\omega\big(q_{\frac{m-n}{2}+1}^{(2,1)}\big)\,,\, \omega\big(q_{\, 3}^{(1,1)}\big)\big\}$
is another weak generating set. 

Unlike the example in Section \ref{Subsec:n+1}, the number of weak generators in $\mathcal{C}^f_{sm}$ can be reduced to two. 
In particular, we prove that $\omega\big(q_{\frac{m-n}{2}+1}^{(1,2)}\big)$ and $\omega\big(q_{\frac{m-n}{2}+1}^{(2,1)}\big)$ weakly generate an element $\omega\big(q_{\, 3}^{(1,1)}\big)$. 
Let $\ell:=\omega\big(q_{\frac{m-n}{2}+1}^{(1,2)}\big)_{(m-n-1)}\omega\big(q_{\frac{m-n}{2}+1}^{(2,1)}\big)$. 
Then using weight 2 elements $\ell$ and ${\ell}_{(1)}\ell$, we obtain elements $\omega\big(q_{\, 2}^{(1,1)}\big)$ and $\omega\big(q_{\, 2}^{(2,2)}\big)$, which, unlike in Section \ref{Subsec:n+1}, cannot be expressed as linear combinations of $\ell$ and ${\ell}_{(1)}\ell$.
Consider a weight three element $\tilde{\ell}:=\omega\big(q_{\frac{m-n}{2}+1}^{(1,2)}\big)_{(m-n-2)}\omega\big(q_{\frac{m-n}{2}+1}^{(2,1)}\big)$.
By computing $\omega\big(q_{\, 2}^{(1,1)}\big)_{(1)}\tilde{\ell}$,
 we have $\omega\big(q_{\, 3}^{(1,1)}\big)$.

\subsection{Rectangular case}
Let $\g$ be $\mathfrak{sl}_N$ or $\mathfrak{sl}_{N_1|N_2}$ with $N_1+N_2=N$ for $N_1\neq N_2$ and $f$ be an even rectangular nilpotent element of $\g$. 
In other words, the Jordan block of $f$ is of type $(m, \cdots, m)$,   where $N_1=d_1\times m\,,\,N_2=d_2\times m$ and $d=d_1+d_2$.
Then by Theorem \ref{big thm} and \ref{big thm super}, 
 the set $\mathcal{C}^f_{\text{big}}=\{\,\omega\big(q_{\, m-1}^{(2,1)}\big)\,,\, \omega\big(q^{(i,i+1)}_{\, m}\big)\,,\, \omega\big(q_{\, m}^{(j+1,j)}\big)\,|\, 1\leq i\leq d-1\,,\, 2\leq j\leq d-1\}
$ weakly generates $\mathcal{W}^k(\g, f)$.

In rectangular case, we can show that the number of weak generators in this case can be reduced to $d$ as follows.
Moreover, we can show that the set 
\begin{align*}
\tilde{\mathcal{C}}^f_{\text{big}}:=\big\{\omega\big(q_{\, m-1}^{(2,1)}\big)\,,\, \omega\big(q_{\, m}^{(1,2)}\big)\,,\,  \omega\big(q_{\, m}^{(i,i+1)}-q_{\, m}^{(i+1,1)}\big)\mid 2\leq i\leq d-1\big\}
\end{align*}
 is a weak generating set. 
In the non-rectangular case, $\omega\big(q_{\frac{m_i+m_{i+1}}{2}}^{(i,i+1)}\big)$ and $\omega\big(q_{\frac{m_1+m_{i+1}}{2}}^{(i+1,1)}\big)$ have different conformal weights, while in the rectangular they have the same conformal weight, making $\tilde{\mathcal{C}}^f_{\text{big}}$ meaningful.
For convenience, assume that $m\geq3$. 
By the proof of Theorem \ref{big thm} and \ref{big thm super}, it suffices to prove that the elements  $
\omega\big(q_{\, m-1}^{(2,1)}\big)$ , $\omega\big(q_{\, m}^{(1,2)}\big)$ and $ \omega\big(q_{\, m}^{(2,3)}-q_{\, m}^{(3,1)}\big)$
weakly generate elements whose linear terms are $ \omega\big(q_{\, m}^{(2,3)}\big)$ and $\omega\big(q_{\, m}^{(3,2)}\big)$, respectively. 
As in Section \ref{Subsec:n+1} and \ref{Subsec:n+2}, we construct  suitable weight two elements and together with $ \omega\big(q_{\, m}^{(2,3)}-q_{\, m}^{(3,1)}\big)$ and their product, obtain the elements with the linear terms $\omega\big(q_{\, m}^{(3,1)}\big)$ and $\omega\big(q_{\, m}^{(2,3)}\big)$.
Then using the $0$-th product of the element whose linear term is $\omega\big(q_{\, m}^{(3,1)}\big)$ and $\omega\big(q_{\, 1}^{(1,2)}\big)$ we have an element with the linear term $\omega\big(q_{\, m}^{(3,2)}\big)$.

On the other hand, by Theorem \ref{big thm small} and \ref{big thm small super}, the set ${\mathcal{C}}^f_{\text{sm}}=\big\{\omega\big(q_{\,3}^{(1,1)}\big)\,,\, \omega\big(q_{\, 2}^{(2,1)}\big)\,,\, \omega\big(q_{\, 1}^{(i, i+1)}\big)\,,\, \omega\big(q_{\, 1}^{(j+1,j)}\big)\mid1\leq i\leq d-1\,, \, 2\leq j\leq d-1\big\}$ is another weak generating set if $m\geq3$. 
In this case too, the number of the weak generating set ${\mathcal{C}}^f_{\text{sm}}$ can be reduced. 
Explicitly, the weak generating set is 
\begin{align*}
\tilde{\mathcal{C}}^f_{\text{sm}}:=\big\{\omega\big(q_{\, 2}^{(2,1)}\big)\,,\,\omega\big(q_{\, 1}^{(1,2)}\big) \, , \, \omega\big(q_{\, 1}^{(i, i+1)}-q_{\, 1}^{(i+1, 1)}\big)\mid2\leq i\leq d-1\big\}.
\end{align*}
As in $\tilde{\mathcal{C}}^f_{\text{big}}$, the set $\tilde{\mathcal{C}}^f_{\text{sm}}$ consists of homogeneous elements.
As shown in the proof that $\tilde{\mathcal{C}}^f_{\text{big}}$ is a weak generating set, we can also show that $\tilde{\mathcal{C}}^f_{\text{sm}}$ weakly generates $\mathcal{W}^k(\g, f)$. 

This result can be related to Theorem 3.17 in \cite{U22}, which deals with a weak generating set of the quantum rectangular case. 
In particular, the author showed that the quantum rectangular W-algebra is weakly generated by two families of elements, $W^{(i, j)}_{\,1}$ and $W^{(i, j)}_{\,2}$,
where $W^{(i, j)}_{\,1}$ has conformal weight $1$, $W^{(i, j)}_{\,2}$ has conformal weight $2$ and the indices $i, j$ range from $1$ to the number of Jordan blocks.  
In other words, if we denote by $d$ the number of Jordan blocks, then the author reduces the number of weak generators to $2d^2\,$.  
The result presented here  reduces the number of weak generators to $d$, the number of Jordan blocks. 

\subsection{Minimal case}
Let $\g$ be $\mathfrak{sl}_N$ or $\mathfrak{sl}_{N_1|N_2}$ with $N_1+N_2=N$ for $N_1\neq N_2$ and $f$ be an even minimal nilpotent element of $\g$. 
Then by Theorem \ref{big thm super}, the weak generating set with large conformal weights is as follows. 
\begin{align*}
\{\omega\big(q_{\ \frac{3}{2}}^{(1, 2)}\big)\,,\, \omega\big(q_{\ \frac{3}{2}}^{(2, 1)}\big)\, , \,\omega\big(q_{\, 1}^{(i,i+1)}\big)\,,\, \omega\big(q_{\, 1}^{(i+1,i)}\big)\mid 2\leq i\leq N-2 \}.
\end{align*}
Using Theorem \ref{big thm small super}, we get the weak generating set with small conformal weights as follows. 
\begin{align*}
\{\omega\big(q_{\ 2}^{(1, 1)}\big)\, , \, \omega\big(q_{\ \frac{3}{2}}^{(1, 2)}\big)\,,\, \omega\big(q_{\ \frac{3}{2}}^{(2, 1)}\big)\, , \,\omega\big(q_{\, 1}^{(i,i+1)}\big)\,,\, \omega\big(q_{\, 1}^{(i+1,i)}\big)\mid 2\leq i\leq N-2 \}.
\end{align*}
Thus, in this case, the weak generating set with large conformal weights obtained using Theorem \ref{big thm super} is the same as the weak generating set with small conformal weights obtained using Theorem \ref{big thm small super}.

\subsection{Example}
Let $\g=\mathfrak{sl}_{13}$ and $f$ be an even nilpotent element whose Jordan block is of type $(6, 4, 3)$. 
Then by Theorem \ref{big thm}, the following set weakly generates $\mathcal{W}^k(\g, f)$.
\begin{align*}
\mathcal{C}^f_{\text{big}}=\big\{A:=\omega\big(q_{\, 5}^{(1,2)}\big)\,,\, B:=\omega\big(q_{\, 5}^{(2,1)}\big)\,,\, C:=\omega\big(q_{\ \frac{7}{2}}^{(2,3)}\big)\,,\, D:=\omega\big(q_{\ \frac{7}{2}}^{(3,2)}\big)\big\}.
\end{align*}
On the other hand, by Theorem \ref{big thm small}, the following set weakly generates $\mathcal{W}^k(\g, f)$.
\begin{align*}
\mathcal{C}^f_{\text{sm}}=\big\{M:=\omega\big(q_{\, 2}^{(1,2)}\big)\,,\, N=\omega\big(q_{\, 2}^{(2,1)}\big)\,,\, P:=\omega\big(q_{\, 3}^{(1,1)}\big)\,,\, R:=\omega\big(q_{\,\frac{3}{2}}^{(2,3)}\big), T:=\omega\big(q_{\,\frac{3}{2}}^{(3,2)}\big)\big\}.
\end{align*}
To compare how to use $n$-th products to get all the stronger generators in Theorem \ref{big thm} and \ref{big thm small}, describe the process of constructing all generators.

First, consider the weak generating set $\mathcal{C}^f_{\text{big}}$. 
Then the elements in $\mathcal{C}^f_{\text{big}}$ weakly generate all the strong generators by using the following  $\lambda$-brackets or products, which are listed below.
\begin{align*}
&\{A\,_\lambda\, B\}\,,\, \{{(A_{(7)}B)}\, _\lambda\, A\}\,,\, \{{(A_{(7)}B)} \,_\lambda\, B\}\,,\, \{X\, _\lambda\, Y\}\,,\, \{C\, _\lambda\, D\}\,,\, Z_{(2)}X\,,\, (Z_{(2)})^2X\,,\, Z_{(2)}Y\,,\, (Z_{(2)})^2Y\,,\,
\\&\{A\,_\lambda\, C\}\,,\, \{B\,_\lambda\, D\}\,,\, W_{(2)}C\,,\, (W_{(2)})^2C\,,\, W_{(2)}D\,,\, (W_{(2)})^2D ,
\end{align*}
where $X=({(A_{(7)}B)}_{(2)}A), Y=({(A_{(7)}B)}_{(2)}B)$ and $Z$ and $W$ are elements whose linear terms are $\omega\big(q_{\, 2}^{(1,1)}\big)$ and $\omega\big(q_{\, 2}^{(2,2)}\big)$, respectively.

Now, consider the weak generating set $\mathcal{C}^f_{\text{sm}}$.
For brevity, we denote by $(a_{(1)})^{n+1}b$ to represent the element obtained by acting with $a_{(1)}$ again on the linear term obtained from $(a_{(1)})^nb$ for $n\geq1$.
Thus, the elements in $\mathcal{C}^f_{\text{sm}}$ weakly generate all the strong generators by using the following $\lambda$-brackets or products, which are listed below.
\begin{align*}
&\{P\,_\lambda\, P\}\, ,\, M_{(0)}R\,,\, N_{(0)}T\,, \, \{M\,_\lambda\, N\}\,,\, \{R\,_\lambda\, T\}\ \,,\,((M_{(1)}N)_{(0)}R)_{(0)}T\,,\, 
\\& U_{(1)}R\,,\, U_{(1)}T\,,\,P_{(1)}M\,,\, P_{(1)}N\,,\, P_{(1)}(M_{(0)}R)\,,\, P_{(1)}(N_{(0)}T)\,,\, (U_{(1)})^2R\,,\,(U_{(1)})^2T, 
\\&(P_{(1)})^2M\,,\,(P_{(1)})^2N\,,\, U_{(1)}U\,,\, (P_{(1)})^2(M_{(0)}R) \,,\,(P_{(1)})^2(N_{(0)}T)\,,\, (P_{(1)})^2P\,,\,(P_{(1)})^3M\,,\,(P_{(1)})^3N\,,\,(P_{(1)})^3P,
\end{align*}
where $U$ is the element whose linear term is $\omega\big(q_{\, 3}^{(2,2)}\big)$. 

Therefore, considering the weak generating set with elements of large conformal weights has the advantage of needing fewer brackets to generate all the strong generators. 
However, it also has the disadvantage of forcing all products to be calculated. 
In contrast, considering the weak generating set with elements small conformal weights takes more brackets, but all strong generators can be generated using just $0$-th and first products after the initial conditions are achieved.


\end{document}